\documentclass[%
  reprint,
  twocolumn,
groupedaddress,
showpacs,preprintnumbers,
 amsmath,amssymb,
 aps,
 prd,
]{revtex4-1}

\usepackage{braket}            
\usepackage{graphicx}          
\usepackage[utf8]{inputenc}    
\usepackage{hyperref}          
\usepackage{dcolumn}           
\usepackage{color}             
\usepackage{adjustbox}         
\usepackage[caption=false]{subfig}        
\usepackage{csquotes}        

\newcommand{\vp}{\vec{p}}
\newcommand{\e}[1]{\text{e}^{#1}}
\newcommand{\vB}{\vec{B}}
\newcommand{\toG}{\Braket{\Omega | T\left\{\chi\left(\vx,t\right)\,\bar{\chi}\left(\vec{0},t\right)\right\}|\Omega}}
\newcommand{\toGr}{\Braket{\Omega | T\left\{\chi\left(r,t\right)\,\bar{\chi}\left(\vec{0},t\right)\right\}|\Omega}}
\newcommand{\pvB}{\psi_{\vB}}

\newcommand{\SUto}{SU(3) $\times$ U(1) }
\newcommand{\abs}[1]{\left|#1\right|}
\newcommand{\order}[1]{\mathcal{O}\left(#1\right)}
\newcommand{\aqeb}{\abs{qe\,B}}
\newcommand{\aeb}{\abs{e\,B}}
\newcommand{\epm}{\left(E+M\right)}

\newcommand{\upup}{{\upharpoonleft\!\upharpoonright}}
\newcommand{\updown}{{\upharpoonleft\!\downharpoonright}}
\newcommand{\vx}{\vec{x}}
\newcommand{\hmu}{\hat{\mu}}
\newcommand{\pivB}{\psi_{i,\vB}}
\newcommand{\ctdof}{\chi^2_{dof}}
\newcommand{\kd}{k_d}
\newcommand{\kB}{k_B}
\newcommand{\vmu}{\vec{\mu}}

\newcommand{\rb}[1]{\left(#1\right)}

\newcommand{\mpi}{m_\pi}
\newcommand{\bpN}{\beta^{M\,B}}
\newcommand{\bpD}{\beta^{M,\,\Bpr}}
\newcommand{\brmpit}{\left(\mpi^2, \Lambda\right)}

\newcommand{\fpi}{f_\pi}
\newcommand{\dthk}{\text{d}^3k}
\newcommand{\vk}{\vec{k}}
\newcommand{\wvk}{\omega_{\vk}}

\newcommand{\mpit}{\mpi^{2}}
\newcommand{\atL}{a_{2}\rb{\Lambda}}
\newcommand{\azL}{a_{0}\rb{\Lambda}}
\newcommand{\Bpr}{B^{\prime}}
\newcommand{\LamFV}{\Lambda^{FV}}

\newcommand{\chit}{\chi^{2}}
\newcommand{\Kp}{{K^{+}}}
\newcommand{\Sigmam}{\Sigma^{-}}
\newcommand{\Sigmap}{\Sigma^{+}}
\newcommand{\Kz}{K^{0}}

\newcommand{\Sigz}{\Sigma^{0}}
\newcommand{\Km}{K^{-}}
\newcommand{\qu}{q_{u}}
\newcommand{\qub}{q_{\overline{u}}}
\newcommand{\qd}{q_{d}}
\newcommand{\qdb}{q_{\overline{d}}}

\newcommand{\qsb}{q_{\overline{s}}}
\newcommand{\Deltaz}{\Delta^{0}}
\newcommand{\SigmamStar}{\Sigma^{*\,-}}
\newcommand{\SigmapStar}{\Sigma^{*\,+}}
\newcommand{\SigzStar}{\Sigma^{*\,0}}
\newcommand{\Deltap}{\Delta^{+}}
\newcommand{\Deltam}{\Delta^{-}}
\newcommand{\pim}{\pi^{-}}
\newcommand{\pip}{{\pi^{+}}}
\newcommand{\Deltapp}{\Delta^{++}}

\newcommand{\piz}{\pi_{0}}
\newcommand{\etaCust}{\eta}
\newcommand{\mcC}{\mathcal{C}}
\newcommand{\Xiz}{\Xi^{0}}
\newcommand{\chiKpSigmamStar}{\chit_{\Kp\,\SigmamStar}}
\newcommand{\chiKzSigzStar}{\chit_{\Kz\,\SigzStar}}
\newcommand{\chiKPSigmamStar}{\chit_{\Kp\,\SigmamStar}}
\newcommand{\chiKpSigzStar}{\chit_{\Kp\,\SigzStar}}
\newcommand{\chiKzSigmapStar}{\chit_{\Kz\,\SigmapStar}}

\newcommand{\chiKpSigmam}{\chit_{\Kp\,\Sigmam}}
\newcommand{\chiKzSigz}{\chit_{\Kz\,\Sigz}}
\newcommand{\chiKpSigz}{\chit_{\Kp\,\Sigz}}
\newcommand{\chiKzSigmap}{\chit_{\Kz\,\Sigmap}}

\newcommand{\chiKzLambda}{\chit_{\Kz\,\Lambda}}
\newcommand{\chiKpLambda}{\chit_{\Kp\,\Lambda}}

\newcommand{\chiDeltappim}{\chit_{\pim\,\Deltap}}
\newcommand{\chiDeltampip}{\chit_{\pip\,\Deltam}}
\newcommand{\chiPipDeltaz}{\chit_{\pip\,\Deltaz}}
\newcommand{\chiPimDeltapp}{\chit_{\pim\,\Deltapp}}

\newcommand{\chinpip}{\chit_{\pip\,n}}

\newcommand{\chiPimp}{\chit_{\pim\,p}}

\newcommand{\eqnrtwo}[2]{Eqs.~$\left(\ref{#1}\right)$ and $\left(\ref{#2}\right)$}
\newcommand{\Fig}[1]{Figure \ref{#1}}
\newcommand{\Tab}[1]{Table \ref{#1}}
\newcommand{\Tabtwo}[2]{Tables \ref{#1} and \ref{#2}}
\newcommand{\Sec}[1]{Section \ref{#1}}
\newcommand{\Figtwo}[2]{Figures \ref{#1} and \ref{#2}}
\newcommand{\Refl}[1]{Ref.~\cite{#1}}    
\newcommand{\eqnr}[1]{Eq.~$\left(\ref{#1}\right)$}

\begin{document}

\preprint{ADP-20-6/T1116}

\title{Magnetic polarisability of the nucleon using a Laplacian mode projection}


\author{Ryan Bignell}
\email[]{ryan.bignell@adelaide.edu.au}
\author{Waseem Kamleh}
\author{Derek Leinweber}
\affiliation{Special Research Centre for the Subatomic Structure of Matter (CSSM),\\
Department of Physics, University of Adelaide, Adelaide, South Australia 5005, Australia
}


\date{\today}

\begin{abstract}
Conventional hadron interpolating fields, which utilise gauge-covariant Gaussian smearing, are ineffective in isolating ground state nucleons in a uniform background magnetic field. There is evidence that residual Landau mode physics remains at the quark level, even when QCD interactions are present. In this work, quark-level projection operators are constructed from the \SUto eigenmodes of the two-dimensional lattice Laplacian operator associated with Landau modes. These quark-level modes are formed from a periodic finite lattice where both the background field and strong interactions are present. Using these eigenmodes, quark-propagator projection operators provide the enhanced hadronic energy-eigenstate isolation necessary for calculation of nucleon energy shifts in a magnetic field. The magnetic polarisability of both the proton and neutron is calculated using this method on the $32^3 \times 64$ dynamical QCD lattices provided by the PACS-CS Collaboration. A chiral effective-field theory analysis is used to connect the lattice QCD results to the physical regime, obtaining magnetic polarisabilities of $\beta^p = 2.79(22)\!\rb{^{+13}_{-18}} \times 10^{-4}$ fm$^3$ and $\beta^n = 2.06(26)\!\rb{^{+15}_{-20}} \times 10^{-4}$ fm$^3$, where the numbers in parentheses describe statistical and systematic uncertainties.
\end{abstract}

\pacs{13.40.-f, 12.38.Gc, 12.39.Fe}

\maketitle

\section{Introduction}
The magnetic polarisability describes the response of a system of charged particles to an external magnetic field. The study of nucleon polarisabilities is an area of key experimental and theoretical interest~\cite{Shanahan:2017bgi,Lensky:2017bwi,Sokhoyan:2016yrc,Mornacchi:2017iuj,Mornacchi:2019rtq,Pasquini:2019nnx}, and the magnetic polarisabilities are key quantities in this area. Measurement of magnetic polarisabilities is difficult~\cite{MacGibbon:1995in,Blanpied:2001ae} and improvement in experimental measurements is evident in recent years~\cite{Tanabashi:2018oca,McGovern:2012ew,Pasquini:2019nnx}. Lattice QCD can play an important role in making predictions in this area.
\par
The uniform background field method has been used successfully to calculate magnetic moments~\cite{Martinelli:1982cb,PhysRevLett.49.1076,Lee:2005ds,Parreno:2016fwu} of hadrons and the magnetic polarisability of neutral particles such as the neutral pion~\cite{Alexandru:2015dva,Bignell:2019vpy} and neutron~\cite{Bignell:2018acn,Chang:2015qxa}. Herein, we present new lattice QCD techniques that enable an investigation of the proton polarisability in an accurate manner.
\par
An uniform external magnetic field is induced through the introduction of an exponential $U(1)$ phase factor on the gauge links across the lattice.
This external field changes the energy of the nucleon according to the energy-field relation~\cite{Martinelli:1982cb,Bernard:1982yu,Burkardt:1996vb,Tiburzi:2012ks,Primer:2013pva,Chang:2015qxa,Bignell:2018acn}
\begin{multline}
  E(B) = m + \vmu \cdot \vB + \rb{2\,n+1}\frac{\aqeb}{2\,m} - \frac{4\,\pi}{2}\,\beta\,B^2 + \ldots 
  \label{eqn:EofB}
\end{multline}
where the nucleon has mass $m$ and magnetic moment and magnetic polarisability $\vmu$ and $\beta$ respectively. The Landau energy term~\cite{QFTZuber} is proportional to $\aqeb$. There is in principle an infinite tower of Landau levels, $\rb{2\,n+1}\,\aqeb/2\,m$ for $n=0,1,2,\dots$ which poses an additional complication for charged hadrons such as the proton. 
\par
While the extraction of the magnetic polarisability seems simple $-$ simply fit the linear and quadratic coefficients of the energies of \eqnr{eqn:EofB} as a function of field strength $-$ this approach is problematic as the magnetic polarisability appears at second-order in the energy of the nucleon~\cite{Burkardt:1996vb,Tiburzi:2012ks,Primer:2013pva,Chang:2015qxa,Bignell:2018acn}. The contribution of the magnetic polarisability to the nucleon energy is necessarily small compared to the overall energy of the particle if the energy expansion of \eqnr{eqn:EofB} is to have small $\order{B^3}$ contributions.
\par
Three-dimensional gauge-covariant Gaussian smearing~\cite{Gusken:1989qx} on the quark fields at the source has been shown to efficiently isolate the nucleon ground state in pure QCD calculations~\cite{Durr:2008zz, Mahbub:2010rm}. This is not the case when a uniform background field is present; the magnetic field breaks three-dimensional spatial symmetry and introduces electromagnetic perturbations into the dynamics of the charged quarks, thus altering the physics present. When QCD interactions are absent, each quark will have a Landau energy proportional to its charge. In the presence of QCD these quarks hadronise, such that in the confining phase the Landau energy will correspond to that of the composite hadron.
\par
It is clear that in the confining phase the effects of the QCD and magnetic interactions compete with each other. Previous studies have demonstrated that Landau physics remains relevant even when QCD interactions are present~\cite{Bruckmann:2017pft,Bignell:2018acn}. This leads us to the idea of using quark operators on the lattice that capture both of these forces. In particular, we have the freedom of choosing asymmetric source and
sink operators in order to construct correlation functions that provide better overlap with the energy eigenstates of the nucleon in a background magnetic field.
\par
The two-dimensional $U(1)$ Laplacian is associated with the Landau modes of a charged particle in a magnetic field. In our previous study of the neutron~\cite{Bignell:2018acn} we considered a quark sink projection based on these two-dimensional $U(1)$ eigenmodes on gauge-fixed QCD fields. In the present study we explore the use of a projector derived from the eigenmodes of the two-dimensional lattice \SUto Laplacian operator. The use of a fully gauge-covariant eigenmode-projected quark sink which encapsulates both QCD and Landau level physics eliminates the need for gauge fixing. We find the use of the \SUto modes as a quark projection operator to be effective in isolating the ground state of the proton in an external magnetic field, enabling an accurate determination of the proton magnetic polarisability.
\par
The presentation of this research is as follows. \Sec{sec:BFM} briefly describes our implementation of a uniform magnetic field. \Sec{sec:MagPol} describes the process by which the magnetic polarisability can be extracted from nucleon two point correlation functions while \Sec{sec:QuarkOps} describes the smeared source and \SUto projected sink used to isolate the nucleon ground states at non-zero magnetic field strengths. The results at several quark masses are presented in \Sec{sec:Results}, and these are used to inform the chiral extrapolations to the physical regime of \Sec{sec:chiEFT}. \Sec{sec:conc} summarises conclusions.

%
\section{Background field method}
\label{sec:BFM}
The following background field method is used to introduce a constant magnetic field along a single axis. This technique is derived first in the continuum where a minimal electromagnetic coupling is added to form the covariant derivative
\begin{align}
	D_\mu = \partial_\mu + i\,qe\,A_\mu,
\end{align}
where $qe$ is the charge of the fermion field and $A_\mu$ is the electromagnetic four potential. On the lattice, the equivalent modification is to multiply the QCD gauge links by an exponential phase factor
\begin{align}
	U_\mu(x) \rightarrow U_\mu(x)\,e^{(i\,a\,qe\,A_\mu(x))}.
\end{align}
As $\vB = \vec{\nabla} \times \vec{A}$, a uniform magnetic field along the $\hat{z}$ axis is obtained via
\begin{align}
  B_z &= \partial_x\,A_y - \partial_y\,A_x.
\end{align}
To give a constant magnetic field of magnitude $B$ in the $+\hat{z}$ direction on the lattice we exploit both $A_x$ and $A_y$. Throughout the lattice we set $A_x = -B\,y$. To maintain the constant magnetic field across the $\hat{y}$ edge of the lattice where periodic boundary conditions are in effect, we set $A_y = + B\,N_y\,x$ along the $\hat{y}$ boundary $y=N_y$. This then induces a quantisation condition for the uniform magnetic field strength~\cite{Primer:2013pva}
\begin{align}
	qe\,B\,a^2 = \frac{2\,\pi\,k}{N_x\,N_y}.
	\label{eqn:qc-q}
\end{align}
Here $a$ is the lattice spacing, $N_x$ and $N_y$ are the spatial dimensions of the lattice, and $k$ an integer specifying the field strength in terms of the minimum field strength.
\par
In this work the field quanta $k$ is in terms of the charge of the down quark, i.e., $k=k_d$ and $q=-1/3$
\begin{align}
	q_de\,B\,a^2 = \frac{2\,\pi\,k_d}{N_x\,N_y}.
	\label{eqn:qc}
\end{align}
Hence a field with $k_d=1$ will be in the $-\hat{z}$ direction. The magnetic field experienced by a baryon is defined to be $\kB = -3\,k_d$.

\section{Magnetic Polarisability}
\label{sec:MagPol}
The naive process of fitting to \eqnr{eqn:EofB} as a function of field strengths is not a viable method with which to extract the magnetic polarisability. Instead a ratio of correlation functions is constructed to isolate the energy shift in a manner enabling correlated QCD fluctuations to be reduced. To form this ratio, we define the spin-field antialigned two-point correlation function
\begin{align}
  G_\updown(B) = G(+s,-B) + G(-s,+B).
  \label{eqn:Gupdown2}
\end{align}
and the spin-field aligned correlator by
\begin{align}
  G_\upup(B) = G(+s,+B) + G(-s,-B).
  \label{eqn:Gupup2}
\end{align}
Here spin-up/down is represented by $\rb{+s/-\!s}$ respectively and the magnetic field orientation along the spin quantisation direction, $\hat{z}$, by $\rb{\pm B}$. These spin-field antialigned and aligned correlators form an improved unbiased estimator for the required correlation functions as they are averages over the required spin and field combinations.
\par
The ratio required to isolate the magnetic polarisability of \eqnr{eqn:EofB} draws on \eqnrtwo{eqn:Gupdown2}{eqn:Gupup2} along with the spin-averaged zero-field correlator $G(0,t)$
\begin{align}
  R(B,t) = \frac{ G_\upup(B,t)\,G_\updown(B,t) }{G(0,t)^2}.
  \label{eqn:R-2}
\end{align}
The zero-field correlator subtracts the mass term from the total energy of the anti-aligned and aligned contributions while the contribution from the magnetic moment term of \eqnr{eqn:EofB} is removed by the product of the spin-field antialigned and aligned correlators. This product yields an exponent of the sum of the aligned and antialigned energy shifts 
\begin{align}
  \frac{\delta E_{\upup}\rb{B} + \delta E_{\updown}\rb{B}}{2} = \frac{\aqeb}{2\,M} - \frac{4\,\pi}{2}\,\beta\,\abs{B}^2 + \order{B^4}.
\end{align}
Thus the desired energy shift is
\begin{align}
  \delta E\rb{B,t} &= \frac{1}{2}\,\frac{1}{\delta\,t}\,\log\rb{\frac{R\rb{B,t}}{R\rb{B,t+\delta\,t}}}
    \label{eqn:logR}\\
&\rightarrow \frac{\aqeb}{2\,M} - \frac{4\,\pi}{2}\,\beta\,\abs{B}^2 + \order{B^4},
    \label{eqn:esubm}
\end{align}
for large Euclidean time. For the neutrally charged neutron, $\aqeb = 0$ and hence the Landau level term vanishes, providing direct access to the polarisability. For the proton this term makes an important contribution and we investigate its magnitude in a variety of fits.
\subsection{Fitting}
For a charged hadron such as the proton, the energy shift for the polarisability is as specified by \eqnr{eqn:esubm} which has a term linear in $B$ and a term quadratic in $B$. As such, an appropriate fit as a function of field strength has both a linear and a quadratic dependence
\begin{align}
  \delta E\rb{\kd} = c_1\,\kd +c_2\,\kd^2
\end{align}
where we fit as a function of $k_d$, the integer magnetic flux quanta in \eqnr{eqn:qc}. $c_1$ and $c_2$ are fit parameters with the units of $\delta E\rb{\kd,t}$. This is in contrast to the neutron where only a quadratic term is required~\cite{Bignell:2018acn}.
\par
As $c_1$ is a free parameter, fitting in this manner allows the charge of the proton to be non-unitary. Thus, we also consider the linear constrained energy shift
\begin{align}
\delta E\rb{\kd} -\frac{\aeb}{2\,M} = c_2\,\kd^2 = -\frac{4\,\pi}{2}\,\beta\,\abs{B}^2 + \order{B^4}.
  \label{eqn:EmmCon}
\end{align}
Here the known linear term has been explicitly subtracted with $q=1$ such that only a term quadratic in $B$ is fitted. This constrains the charge of the proton to $q_p=1$.
\par
The quantisation condition of \eqnr{eqn:qc} provides
\begin{align}
  \beta = -2\,c_2\,\alpha\,q_d^2\,a^4\,\rb{\frac{N_x\,N_y}{2\,\pi}}^2,
  \label{eqn:betaConv}
\end{align}
where $\alpha = 1/137\dots$ is the fine structure constant.
\par
The energy shifts which are fit in \eqnr{eqn:EmmCon} must be well determined at all field strengths. We apply the single state ansatz, requiring that a constant plateau fit can be found as a function of Euclidean lattice time $t$. Correlations between adjacent Euclidean time slices are considered through the use of the full covariance matrix $\ctdof$ which is estimated via the jackknife method~\cite{Efron1979}. The resulting fit as a function of field strength must then also fit the energy shifts with an acceptable $\chi^2$ per degree of freedom.
\par
\par
Fit windows were kept consistent across the three non-zero field strengths considered where possible. Where this proved difficult, the fit windows between field strengths were allowed to vary in a monotonic manner, with the lowest field strength having the longest fit region. Through this process we ensure that the lowest lying state is isolated for each energy shift.


%


\section{Quark Operators}
\label{sec:QuarkOps}
The use of asymmetric source and sink operators enable the construction of nucleon correlation functions which have greater overlap with the lowest energy eigenstates of the nucleon in a background magnetic field. This is important as the energy shift required for the magnetic polarisability is small compared to the total energy. The signal becomes disguised by noise at late Euclidean time.
\par
The quark source is constructed using three-dimensional gauge-invariant Gaussian smearing~\cite{Gusken:1989qx} as is common practice in lattice QCD~\cite{Durr:2008zz,Mahbub:2010rm} while the quark sink uses the \SUto eigenmode quark-projection method discussed below.

\subsection{\SUto eigenmode projection}
\label{sec:suto}

For a charged particle in a uniform magnetic field the
lattice Landau levels are eigenmodes of the  (2D) $U(1)$ Laplacian. 
Here we wish to construct a fully gauge-covariant quark sink projection operator that encompasses QCD as well as the electromagnetic potential.
To do so, we calculate the low-lying eigenmodes $\ket{\psi_i}$ of the two-dimensional lattice Laplacian
\begin{align}
  \Delta_{\vx,\vx^\prime} = 4\,\delta_{\vx,\vx^\prime} -\! \sum_{\mu=1,2}U_\mu(\vx)\delta_{\vx+\hmu,\vx^\prime} + U^\dagger_\mu(\vx-\hmu)\delta_{\vx-\hmu,\vx^\prime},
  \label{eqn:2DLap}
\end{align}
where $U_\mu\left(\vx\right)$ are the full \SUto gauge links as applied in the full lattice QCD calculation. We can then define a projection operator by truncating the completeness relation
\begin{align}
1 = \sum_{i=1}\,\ket{\psi_i}\bra{\psi_i}.
\end{align}
In the pure $U(1)$ case, the lowest Landau level on the lattice has a degeneracy equal to the magnetic flux quanta $|k|$ given in Eq.~\ref{eqn:qc-q}, providing a natural place to truncate the above sum.  The introduction of the QCD interactions into the Laplacian causes the $U(1)$ modes associated with the different Landau levels to mix, such that it is no longer possible to clearly identify the modes associated with the lowest Landau level at small field strengths. Instead, we simply choose a fixed number $n > |k|$ modes to project. This truncation has a similar effect to performing (2D) smearing, by filtering out the high frequency modes. Indeed, we find that for small values of $n$ the projected hadron correlator becomes noisy, just as it does when performing large amounts of sink smearing.
\begin{figure}
  \includegraphics[width=\columnwidth]{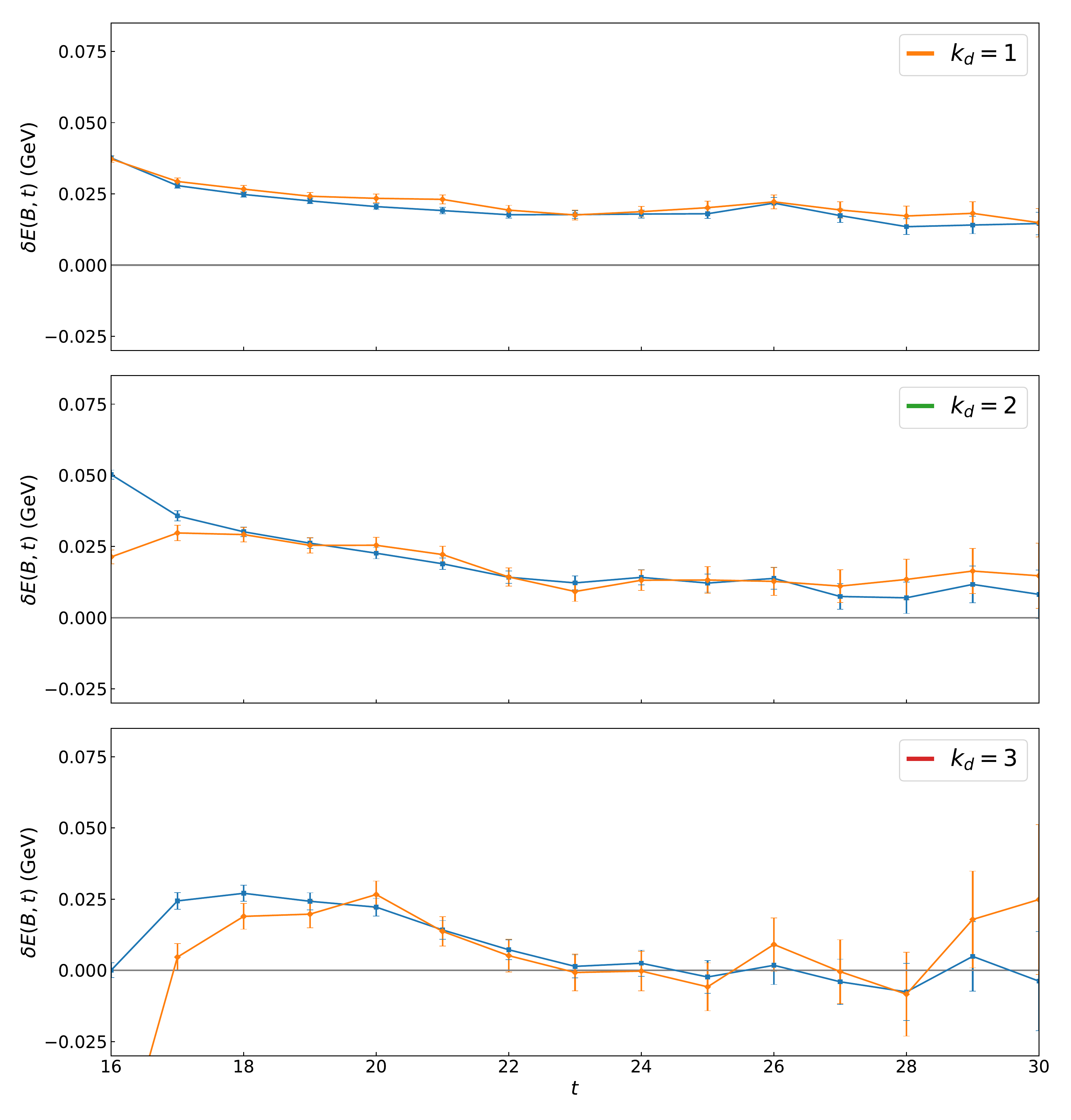}%
  \caption{\label{fig:SUto:LP64-96}The proton energy shift $\delta E\rb{\kd, t}$ of \eqnr{eqn:logR} for three field strengths using $64$ (orange diamond) and $96$ (blue square) eigenmodes in the projection operator of \eqnr{eqn:coordP_n}. The $\mpi = 0.702$ GeV ensemble is shown.}
\end{figure}
\par
The eigenmode truncation is chosen to be sufficiently large so as to avoid introducing large amounts of noise into the two-point correlation function, but also small enough to place a focus on the low-energy physics relevant to isolating the magnetic polarisability. Truncation at $32$, $64$ and $96$ modes are investigated in a manner similar to that at the source. While $32$ modes was not effective, it can be observed in \Fig{fig:SUto:LP64-96} that both $64$ and $96$ eigenmodes produce consistent behaviour in the proton energy shift.
\par
Due to the two-dimensional nature of the Laplacian, the low-lying eigenspace is calculated independently for each $(z,t)$-slice on the lattice. Consequently, we can interpret the four-dimensional coordinate space representation of an eigenmode
\begin{align}
  \braket{\vx,t\,|\,\pivB} = \pivB(x,y\,|\,z,t),
  \label{eqn:coordpsi}
\end{align}
as selecting the two-dimensional coordinate space representation $\pivB(x,y)$ from the eigenspace belonging to the corresponding $(z,t)$-slice of the lattice. The four-dimensional coordinate space representation of the projection operator follows,
\begin{align}
  P_n\left(\vx,t;\vx^\prime,t'\right) = \sum_{i=1}^{n}\,\braket{\vx,t\,|\,\pivB}\braket{\pivB\,|\,\vx^\prime,t'}\,\delta_{zz'}\,\delta_{tt'}.
  \label{eqn:coordP_n}
\end{align}
The Kronecker delta functions in the definition above ensure that the outer product is only taken between eigenmodes from the same subspace (i.e. the projector acts trivially on the $(z,t)$ coordinates).
\par
This projection operator is then applied at the sink to the quark propagator in a coordinate-space representation as
\begin{align}
  S_n\left(\vx,t;\vec{0},0\right) = \sum_{\vx^\prime}\,P_n\left(\vx,t;\vx^\prime,t\right)\,S\left(\vx^\prime,t;\vec{0},0\right).
\end{align}
We select $n=96$ modes for our analysis.
\par
Using the \SUto eigenmode quark-projection operator and a tuned smeared source produces nucleon correlation functions at non-trivial field strengths where the proton is in the QCD ground state and the $n=0$ lowest lying Landau level approximation is justified. The energy shifts required by \eqnrtwo{eqn:logR}{eqn:esubm} display good plateau behaviour, as exhibited in \Figtwo{fig:SUto:LP64-96}{fig:k2dEpol}.

\subsection{Source Smearing}
Whilst we attempt to encapsulate the quark-level physics of the electromagnetic interaction at the sink, we use a smeared source to provide a representation of the QCD interactions with the intent of isolating the QCD ground state. A broad range of smearing levels are examined at zero field strength, $B=0$ in order to do this.
\par
\begin{figure}
  \includegraphics[width=\columnwidth]{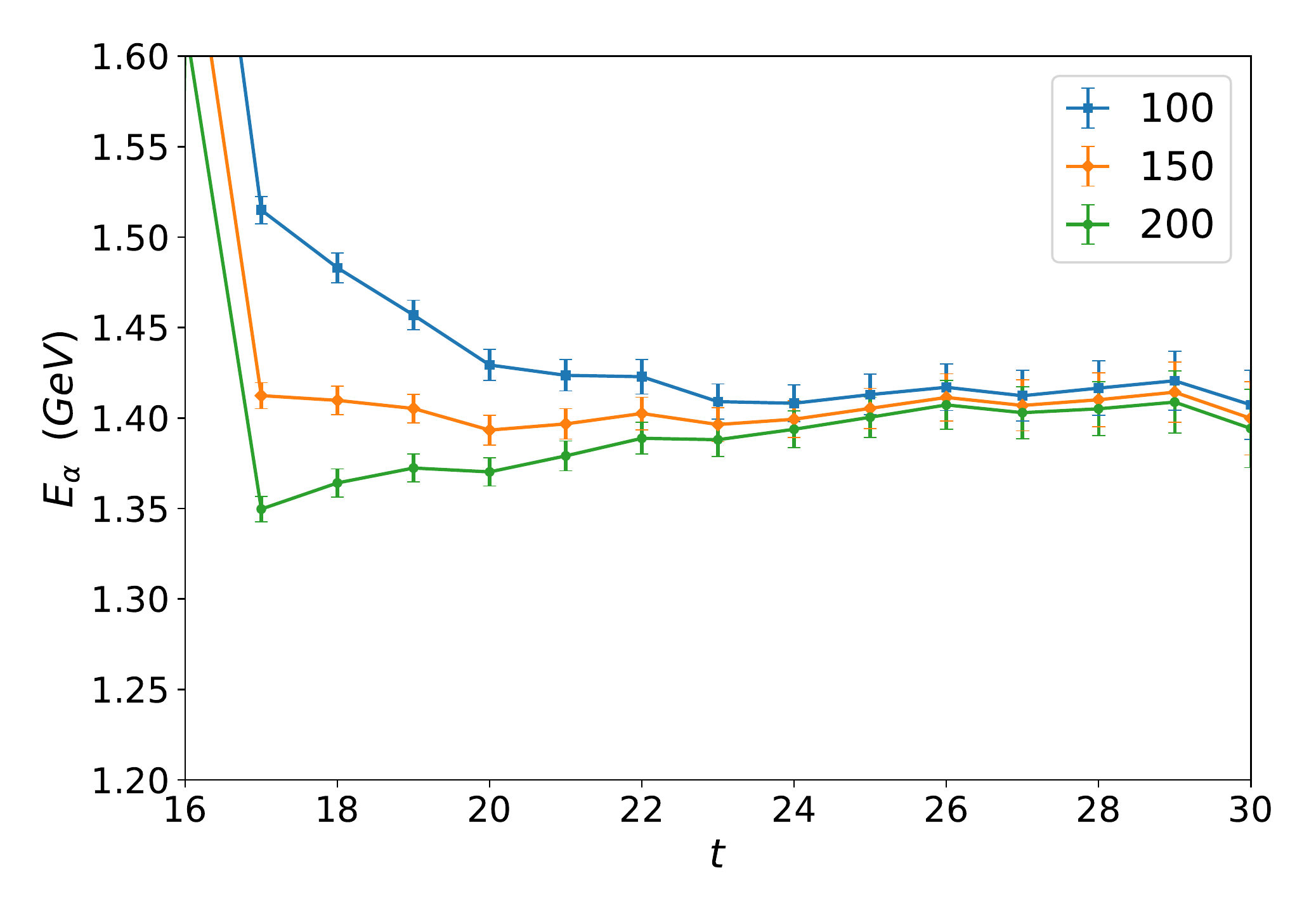}%
  \caption{\label{fig:ManySourcek13700}Proton zero-field effective masses from smeared source to \SUto eigenmode projected sink correlators using various levels of covariant Gaussian smearing at the source on the $\mpi = 0.702$ GeV ensemble. The source is at $t=16$.}
\end{figure}
\begin{figure}
  \includegraphics[width=\columnwidth]{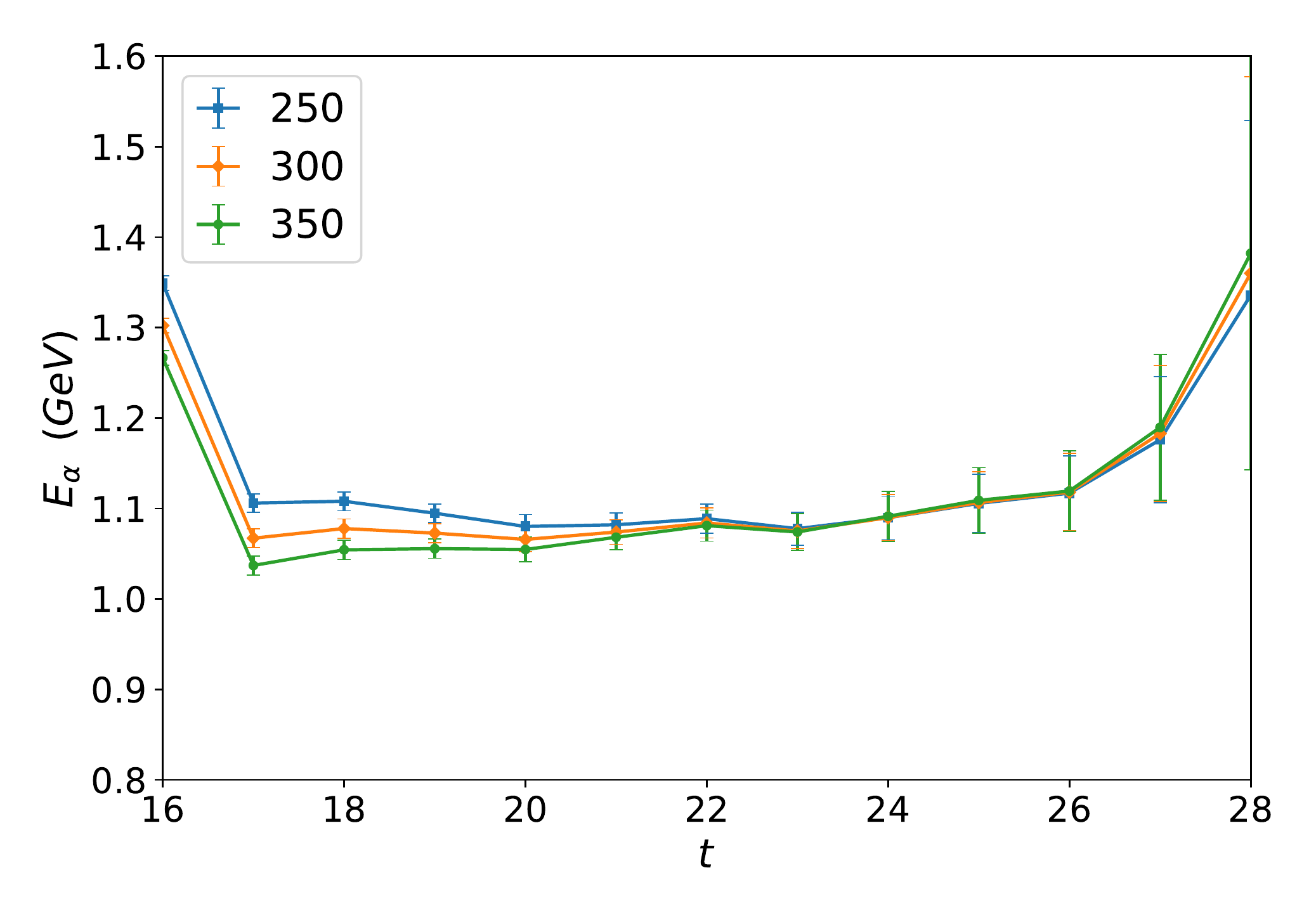}%
  \caption{\label{fig:ManySourcek13770BF0}Proton zero-field effective masses from smeared source to \SUto eigenmode projected sink correlators using various levels of covariant Gaussian smearing at the source on the $\mpi = 0.296$ GeV ensemble. The source is at $t=16$.}
\end{figure}

The effective mass at $B=0$ was investigated for each ensemble and the smearing which produces the earliest onset of plateau behaviour is chosen. This $B=0$ effective mass is shown in \Fig{fig:ManySourcek13700} for $\mpi = 0.702$ GeV where the optimal smearing of $150$ sweeps is chosen. On the lightest ensemble considered, at $\mpi = 0.296$ GeV as shown in \Fig{fig:ManySourcek13770BF0}, the choice is not as obvious. In this case the full set of correlation functions at each finite field-strength is run for each smearing and these results examined.
\begin{figure}
  \includegraphics[width=\columnwidth]{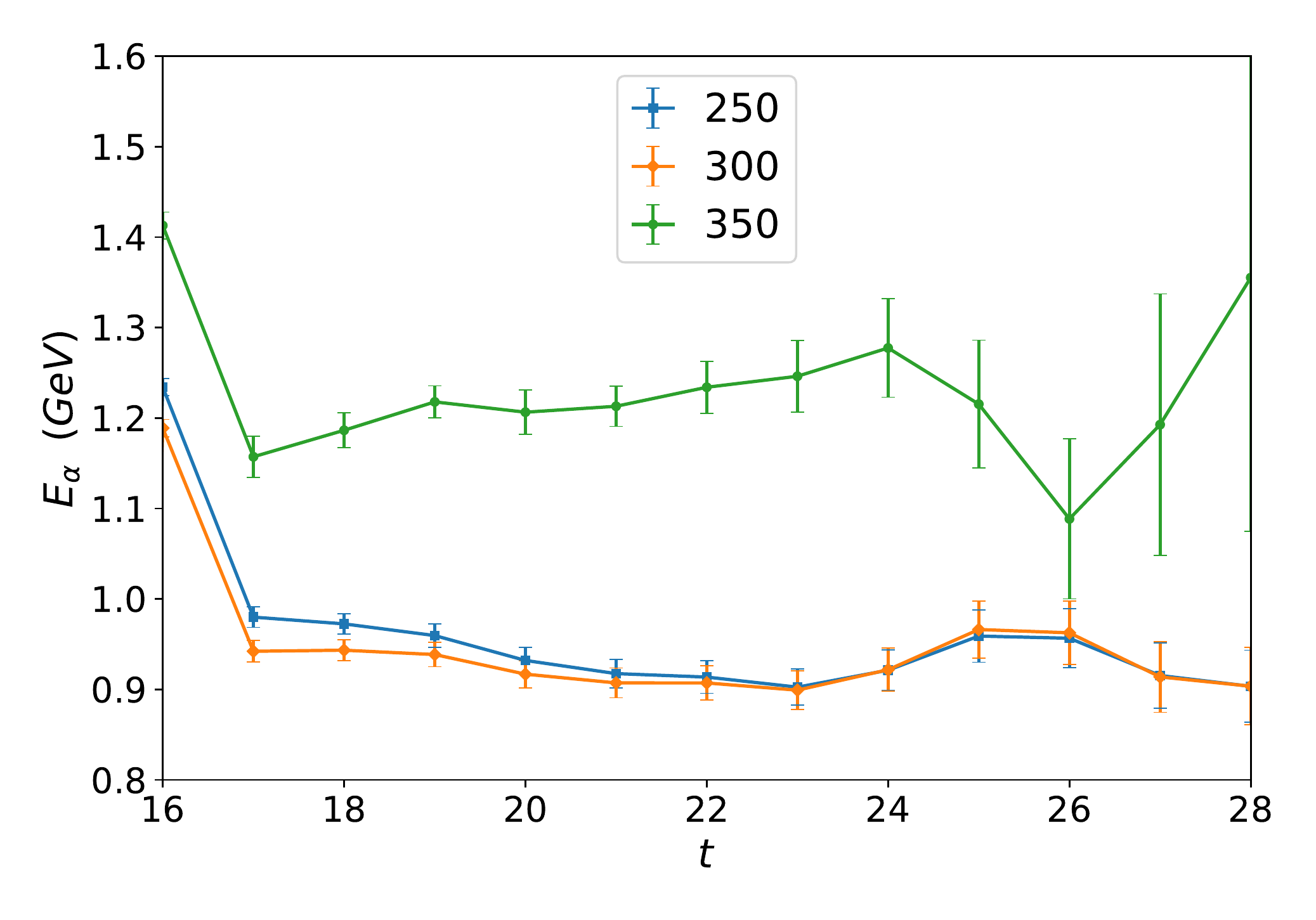}%
  \caption{\label{fig:ManySourcek13770BF2AA}Proton $\kd=2$ anti-aligned effective energies from smeared source to \SUto eigenmode projected sink correlators using various levels of covariant Gaussian smearing at the source on the $\mpi = 0.296$ GeV ensemble. The source is at $t=16$.}
\end{figure}
\par
This reveals a particularly interesting problem with large amounts of smearing which can be seen in the anti-aligned energy shown in \Fig{fig:ManySourcek13770BF2AA}. The anti-aligned energy is examined in preparation its use in the energy shift ratio of \eqnr{eqn:R-2} and has spin and magnetic field anti-aligned as in \eqnr{eqn:Gupdown2}.
\par
When the source is excessively smeared, the larger field strengths couple preferentially to higher Landau levels rather than the lowest. This is evident in how the $350$ sweeps effective energy differs from the other smearings in both value and slope. This difference is close to the difference between Landau levels for the proton, i.e. for
\begin{align}
  E_n^2\rb{B} \sim m^2 + \abs{e\,B}\,\rb{2\,n+1},
\end{align}
the difference can be determined by considering the relativistic energy difference $E^2_{1}\rb{B} - E^2_{0}\rb{B} = 2\,\abs{e\,B}.$ The difference of squares can be factored as
\begin{align}
  \rb{E_{1}\rb{B} - E_{0}\rb{B}}\,\rb{E_{1}\rb{B} + E_{0}\rb{B}} &= 2\,\abs{e\,B}.
\end{align}
Defining $\Delta\,E_{10}\rb{B} = E_{1}\rb{B} - E_{0}\rb{B}$ as the energy difference visible in \Fig{fig:ManySourcek13770BF2AA}, we obtain a quadratic form
\begin{align}
  \Delta\,E_{10}\rb{B}\,\rb{ \Delta\,E_{10}\rb{B} + 2\,E_{0}\rb{B}} - 2\,\abs{e\,B} = 0.
\end{align}
Recalling that the field strength experienced by the proton is related to that of the down quark by $k_B = -3\,\kd$, the appropriate values for \Fig{fig:ManySourcek13770BF2AA} are
\begin{align}
  E_{0}\rb{\kd=2} &\sim 0.9 \text{ GeV}\nonumber \\
  \abs{e\,B\rb{\kd=2}} &\sim 0.522 \text{ GeV}^2,\nonumber
\end{align}
and the energy difference between these two Landau levels is $\Delta\,E_{10}\rb{\kd=2} \sim 0.46$ GeV. This is consistent with the difference between smearings visible in \Fig{fig:ManySourcek13770BF2AA}.
\par
\par
The smeared source examination is followed at each of the quark masses where for masses $m_\pi = 702$, $570$, $411$, $296$ MeV, optimal smearings of $N_{sm} = 150$, $175$, $300$, $250$ respectively are obtained.
\par
An advantage of using the $U(1)\times SU(3)$ Laplacian projector is that it is well defined at zero magnetic field strength, where the $U(1)$ field is equal to unity. This means that the fluctuations at finite $B$ and $B=0$ are strongly correlated, such that they cancel out when taking the ratio of the correlators in \eqnr{eqn:R-2}, providing an improved signal in comparison to the $U(1)$ projection. This improvement does come at a computational cost, as the $U(1)\times S\!U(3)$ Laplacian eigenmodes must be calculated on every configuration. Using the \SUto eigenmode quark-projection operator and a tuned smeared source produces nucleon correlation functions at non-trivial field strengths where the proton is in the QCD ground state and the $n=0$ lowest lying Landau level approximation is justified. This is demonstrated by the energy shifts required by \eqnrtwo{eqn:logR}{eqn:esubm}, which display good plateau behaviour as exhibited in \Figtwo{fig:SUto:LP64-96}{fig:k2dEpol}.
\subsection{Hadronic Landau Projection}
\begin{figure}[t]
        \centering
	\includegraphics[width=0.475\columnwidth,page=1]{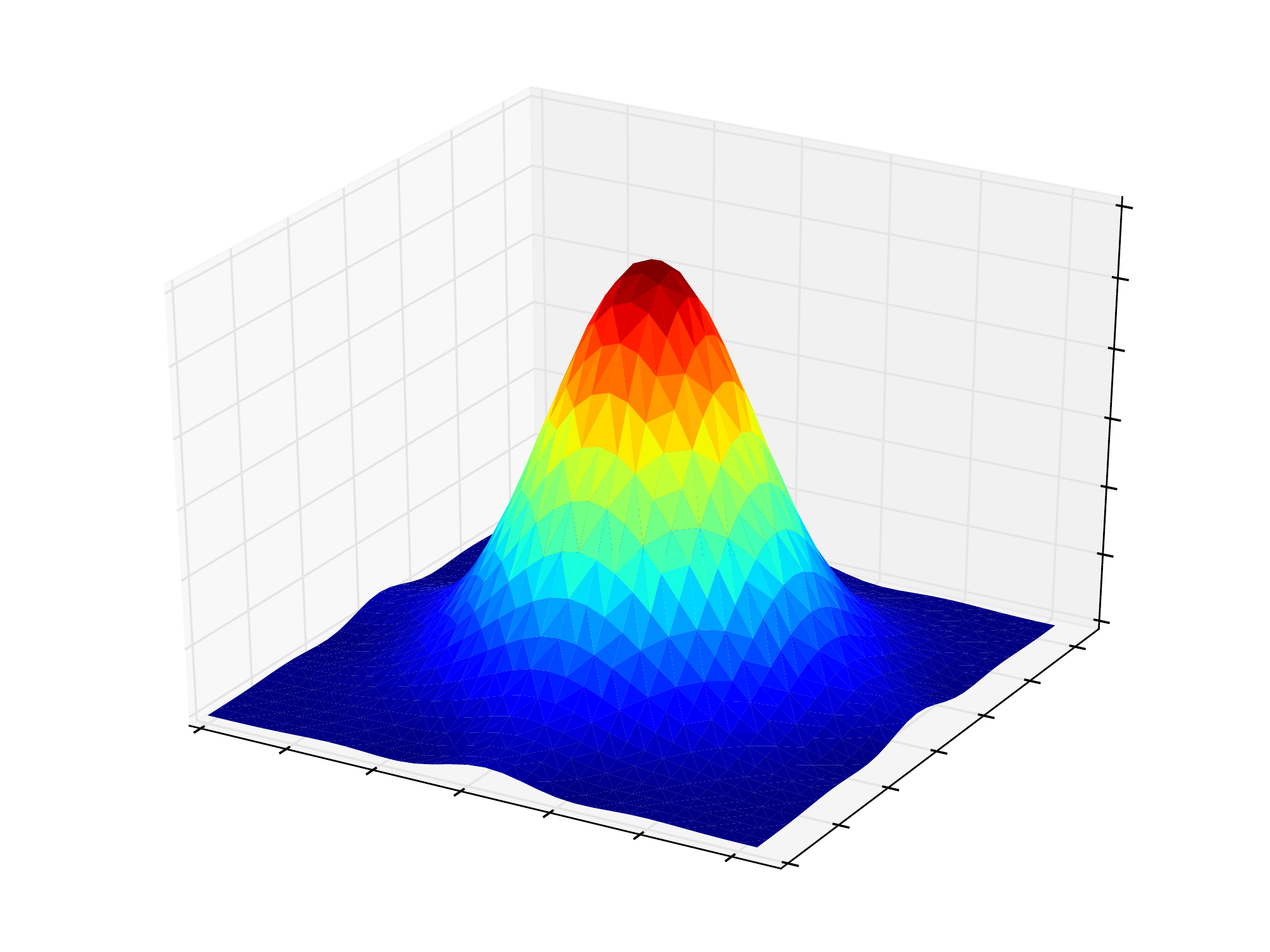}
	\includegraphics[width=0.475\columnwidth,page=2]{ShiftLandauViewNewReSRCnx32ny32kB-6.pdf}
        \\
	\includegraphics[width=0.475\columnwidth,page=3]{ShiftLandauViewNewReSRCnx32ny32kB-6.pdf}
        \includegraphics[width=0.475\columnwidth,page=4]{ShiftLandauViewNewReSRCnx32ny32kB-6.pdf}
        \\
	\includegraphics[width=0.475\columnwidth,page=5]{ShiftLandauViewNewReSRCnx32ny32kB-6.pdf}
        \includegraphics[width=0.475\columnwidth,page=6]{ShiftLandauViewNewReSRCnx32ny32kB-6.pdf}
        \caption{Lowest-lying U(1) eigenmode probability densities of the lattice Laplacian operator in a constant background magnetic field oriented in the $\hat{z}$ direction are plotted as a function of the $x,y$ coordinates. As $\kB = -3\,\kd$, the degenerate eigenmodes for the sixth quantised field strength relevant to the proton for $\kd=2$ are displayed in a linear combination that maximises the overlap of the first mode with the source. The origin is at the centre of the $x-y$ plane.}
	\label{fig:EigenmodePlots}
\end{figure}
The \SUto eigenmode projection technique defined above is relevant to Landau effects at the quark-level. As the proton is a charged hadron it will also experience Landau level behaviour in a magnetic field. The Landau-level physics at the hadronic level is easier to capture due to the colour-singlet nature of the proton. We can simply use the eigenmodes of the $U(1)$ Laplacian, with a well defined degeneracy for the lowest Landau level~\cite{Kamleh:2017yjx,Bignell:2018acn}. Below we describe in detail the prescription for the hadronic Landau level projection.
\par
To study hadronic two-point correlation functions in the zero-field case one calculates the momentum-projected correlator
\begin{align}
  G\left(\vp,t\right) = \sum_{\vx}\,\e{-i\,\vp\cdot\vx}\,\toG,
\end{align}
where $\chi$ and $\bar{\chi}$ are appropriate interpolating fields.
\par
This standard approach of a three-dimensional Fourier projection is not appropriate for the proton when the uniform background magnetic field is present. The presence of the background field causes the energy eigenstates of the charged proton to no longer be eigenstates of the $p_x, p_y$ momentum components. Hence, we instead project the $x,y$ dependence of the two-point correlator onto the lowest Landau level, $\psi_{\vB}\rb{x,y}$, explicitly, and also select a specific value for the $z$ component of momentum,
\begin{align}
  G\left(p_z, \vB,t\right) = \sum_{r}\,&\pvB\left(x,y\right)\,\e{-i\,p_z\,z}\nonumber \\
  &\times \toGr.
\end{align}
In the continuum limit, the lowest Landau mode has a Gaussian form, $\pvB \sim \e{-\aqeb\,\left(x^2+y^2\right)/4}$. However, in a finite volume the periodicity of the lattice causes the wave function's form to be altered~\cite{Tiburzi:2012ks,Bignell:2017lnd}. As such, we instead calculate the lattice Landau eigenmodes using the two-dimensional U(1) gauge-covariant lattice Laplacian in an analogous way to \eqnr{eqn:2DLap}~\cite{Kamleh:2017yjx,Bignell:2018acn}. Here $U_\mu$ contains only the $U(1)$ phases appropriate to the background magnetic field quantised on the lattice.
\par
The correlator projection is then onto the space spanned by the degenerate modes $\pivB$ associated with the lowest lattice Landau level available to the proton
\begin{align}
  G\left(p_z, \vB,t\right) = \sum_{r}\,\sum_{i=1}^{n}\,&\pivB\left(x,y\right)\,\e{-i\,p_z\,z}\nonumber \\
  &\times \toGr.
  \label{eqn:19Draft}
\end{align}
The degeneracy of the lowest-lying Landau mode is given by the magnetic-field quanta $\abs{\kd}$.
\par
In evaluating \eqnr{eqn:19Draft}, we also consider the case of fixing $n=1$, such that only the first eigenmode having the best overlap with the source is considered. Assuming the source $\rho(x,y)=\delta_{x,0}\delta_{y,0}$ is located at the origin, the overlap with the first mode $i=1$ is optimised through a rotation of the $U(1)$ eigenmode basis that maximises the value of $|\braket{\rho\,|\,\psi_{i=1,\vec{B}}}|^2$. An optional phase can be applied so that $\psi_{i=1,\vec{B}}(0,0)$ is purely real at the source point.
\par
In most cases the results are almost indistinguishable and we proceed with $n=\abs{3\,\kd}$. The only exception is for the ensemble with $\mpi = 0.411$ GeV. Here the $i=1$ mode alone provides superior results. As discussed in \Sec{sec:BFM}, for the proton $\kB = -3\,\kd$ and therefore the degeneracy is $n=\abs{3\,\kd}$. \Fig{fig:EigenmodePlots} illustrates the six degenerate modes associated with $\kd=2$.
\par
More generally, this hadronic eigenmode-projected correlator offers superior isolation of the ground state for the proton~\cite{Tiburzi:2012ks} and is crucial for the identification of constant plateaus in the energy shift of \eqnr{eqn:logR}.

\subsection{Simulation Details}
\begin{table*}[!]
  \caption{Lattice simulation parameters with corresponding statistics used.}
  \label{tab:simdet}
  \begin{ruledtabular}
    \begin{tabular}{lllllll}
      $\kappa_{ud}$ & a (fm)          & $\mpi$ (GeV) & $L^3 \times T$   & $\mpi\,L$ & N$_{\text{src}}$                         \\
      0.13700       & 0.1023          & 0.702         & $32^3 \times 64$ & 11.6       & 5 \\
      0.13727       & 0.1009          & 0.570         & $32^3 \times 64$ & 9.4        & 4 \\
      0.13754       & 0.0961          & 0.411         & $32^3 \times 64$ & 6.4        & 4 \\
      0.13770       & 0.0951          & 0.296         & $32^3 \times 64$ & 4.5        & 7 
    \end{tabular}
  \end{ruledtabular}
\end{table*}

The $2+1$ flavour dynamical gauge configurations provided by the PACS-CS~\cite{Aoki:2008sm} collaboration through the ILDG~\cite{Beckett:2009cb} are used in this work. These configurations span a variety of masses, allowing a chiral extrapolation to be performed. A non-perturbatively improved clover fermion action and Iwasaki gauge action provide a physical lattice spacing of $a = 0.0907(13)$ fm. Four different ensembles are considered, corresponding to four different pion masses. The lattice spacing for each ensemble was set using the Sommer scale with $r_0 = 0.49$ fm. The details of each of these ensembles, including the pion mass and statistics used can be found in \Tab{tab:simdet}. Fixed boundary conditions in the time direction are used and the source placed at $N_t/4 = 16$. Source locations are then systematically varied to produce large distances between adjacent sources~\cite{Bignell:2018acn}.
\par
We calculate correlation functions at four distinct magnetic-field strengths - including zero. Propagators at ten nonzero field-strengths are calculated to achieve this at {$e\,B = \pm 0.087$, $\pm 0.174$, $\pm 0.261$, $\pm 0.348$, $\pm 0.522$ GeV$^2$}; corresponding to {$k_d = \pm 1, \, \pm 2,\, \pm 3,\,\pm 4,\,\pm 6$} in \eqnr{eqn:qc}.
\par
The additive mass renormalisation due to the Wilson term~\cite{Bali:2015vua,Bali:2017ian} is removed through use of the Background-Field-Corrected-Clover action~\cite{Bignell:2019vpy}. As the background field is known analytically; a tree-level contribution of $F^B_{\mu\nu}$ from the background field can be included in the clover term of the fermion action, avoiding the non-perturbative improvement coefficient, $C_{SW}$. This removes the non-physical magnetic-field induced additive mass renormalisation due to the Wilson term of the fermion action.
\par
The configurations used are electro-quenched; the magnetic field exists only for the valence quarks of the hadron. While it is possible to include the background field on each configuration~\cite{Fiebig:1988en} this requires a separate Monte Carlo simulation for each field strength and so is prohibitively expensive. Performing separate calculations would also remove the correlated QCD fluctuations between finite-field and zero-field correlation functions, reducing the efficacy of \eqnr{eqn:R-2}.

\section{Results}
\label{sec:Results}
The formalism for extracting the magnetic polarisability of the nucleons using lattice QCD and the background field method has now been established. Hadronic Landau level projected correlation functions are computed at several non-zero field strengths using a specialised \SUto quark sink. Thus hadronic as well as quark level Landau energy level effects are considered in order to isolate the ground state energy of the nucleon in an external magnetic field. A tuned smeared source provides a good representation of the QCD ground state.
\par
\begin{figure}
  \includegraphics[width=\columnwidth]{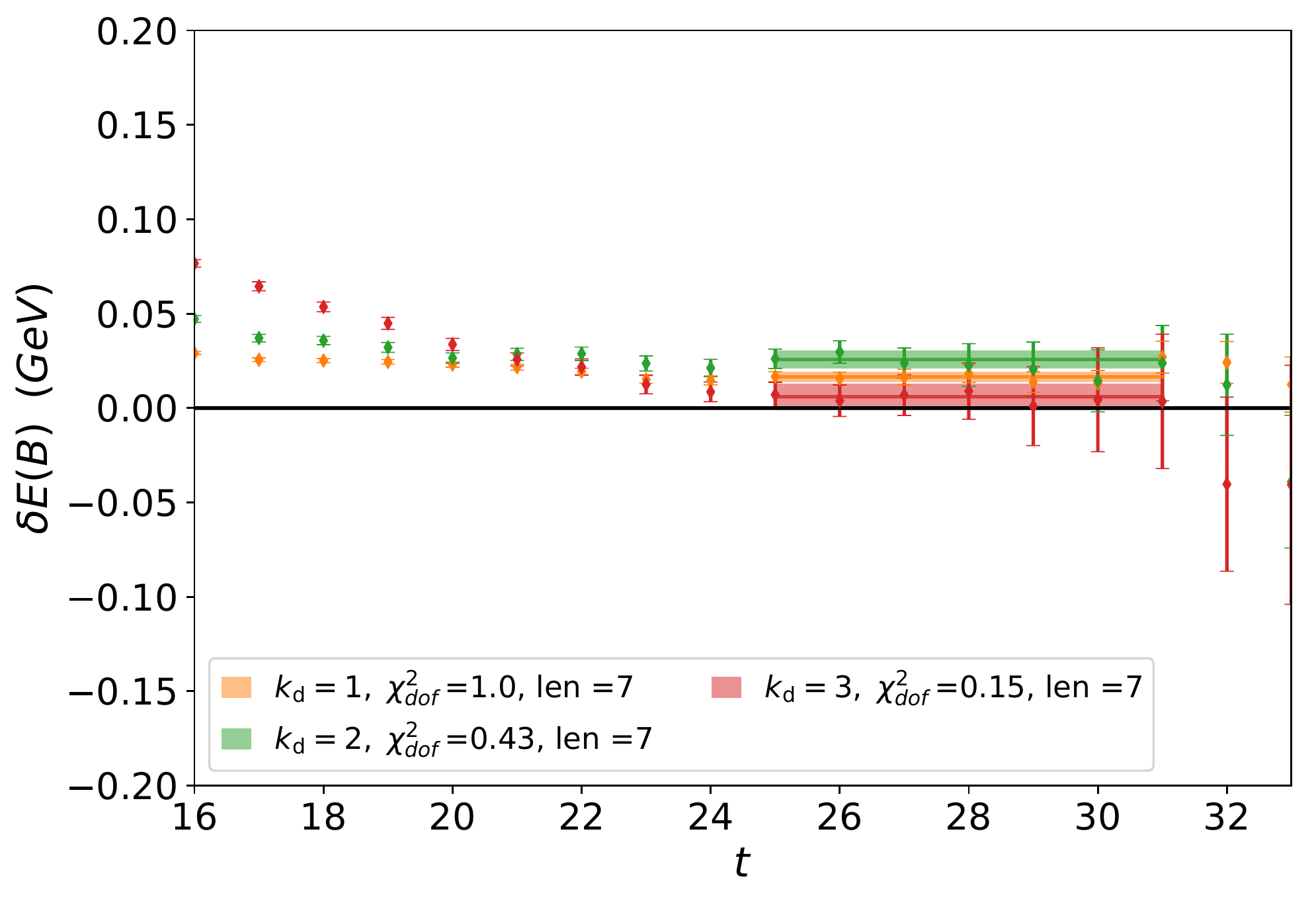}%
  \caption{\label{fig:k2dEpol}The magnetic polarisability effective energy shift, $\delta E\rb{B,t}$ of \eqnr{eqn:logR} for the $\mpi=0.570$ GeV proton as a function of Euclidean time (in lattice units), using a smeared source and the \SUto quark-eigenmode projection technique. Results for field strengths $\kd=1,2,3$ are shown. The selected fits for this ensemble and the $\chi^2_{dof}$ are also illustrated.}
\end{figure}
\begin{figure}
  \includegraphics[width=\columnwidth]{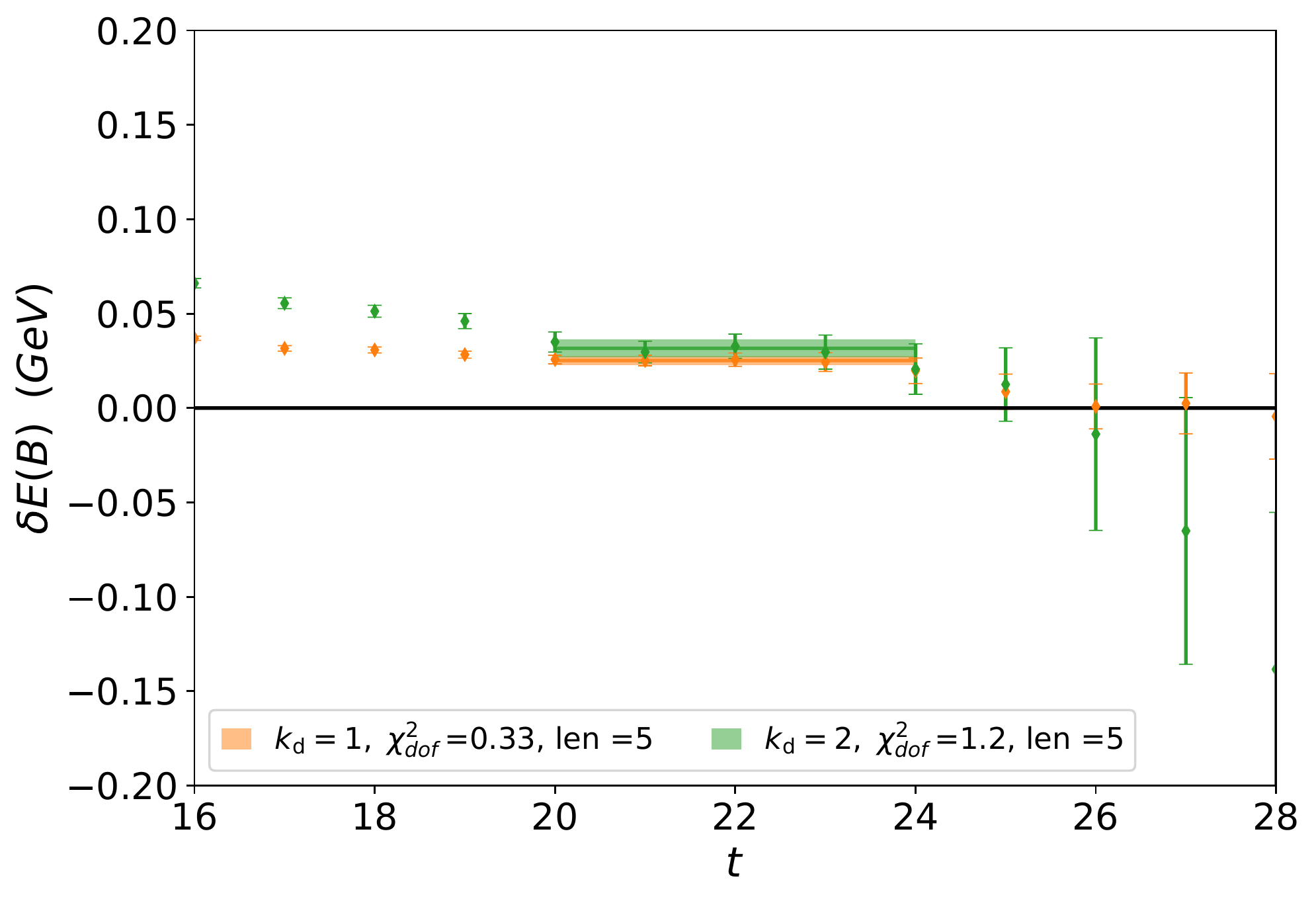}%
  \caption{\label{fig:k4dEpol}The magnetic polarisability effective energy shift, $\delta E\rb{B,t}$ of \eqnr{eqn:logR} for the $\mpi=0.296$ GeV proton as a function of Euclidean time (in lattice units), using a smeared source and the \SUto quark-eigenmode projection technique. Results for field strengths $\kd=1$ and $2$ are shown. The selected fits for this ensemble and the $\chi^2_{dof}$ are also illustrated.}
\end{figure}
The effectiveness of this approach is visible in \Figtwo{fig:k2dEpol}{fig:k4dEpol} where the energy shift required to extract the magnetic polarisability of the proton is plotted for the $\mpi = 0.570$ GeV and $\mpi = 0.296$ GeV ensembles respectively. The effective energy shifts display good plateau behaviour across all three non-zero field strengths. This is a common feature across the heavier three quark masses considered. Good plateaus can be found at all three field strengths.
\par
At the lightest quark mass considered, the third field strength does not present good plateau behaviour. As such only the first two field strengths are considered.
\par
The fit function of \eqnr{eqn:EmmCon} is applied to the non-zero energy shifts produced by \eqnr{eqn:logR}. The constant Euclidean time fits to \eqnr{eqn:logR} are selected using a strict $\ctdof$ criteria, where we require $\ctdof \le 1.2$. This ensures single-state dominance in the energy shift.
\par
In determining the optimal Euclidean-time fit windows to use, each possible fit window across all three field strengths is considered. Where all the Euclidean-time plateau and field-strength dependent fits are acceptable, magnetic polarisability values are calculated from the quadratic coefficient $c_2$ of \eqnr{eqn:EmmCon}.
\par
Where this fitting process yields no or only a small number of acceptable fit windows, we also allow the Euclidean-time fit window at each field strength to vary. These fits must still have a common fit end point but the fit start point is allowed to increase in a monotonic manner with increasing field strength. The smallest field strength must have the longest fit window, the second field strength the second longest and similarly for further field strengths. This fitting process expands the fit window parameter space available and is particularly helpful at lighter quark masses.
\begin{figure}[t]
        \centering
	\includegraphics[width=0.475\columnwidth]{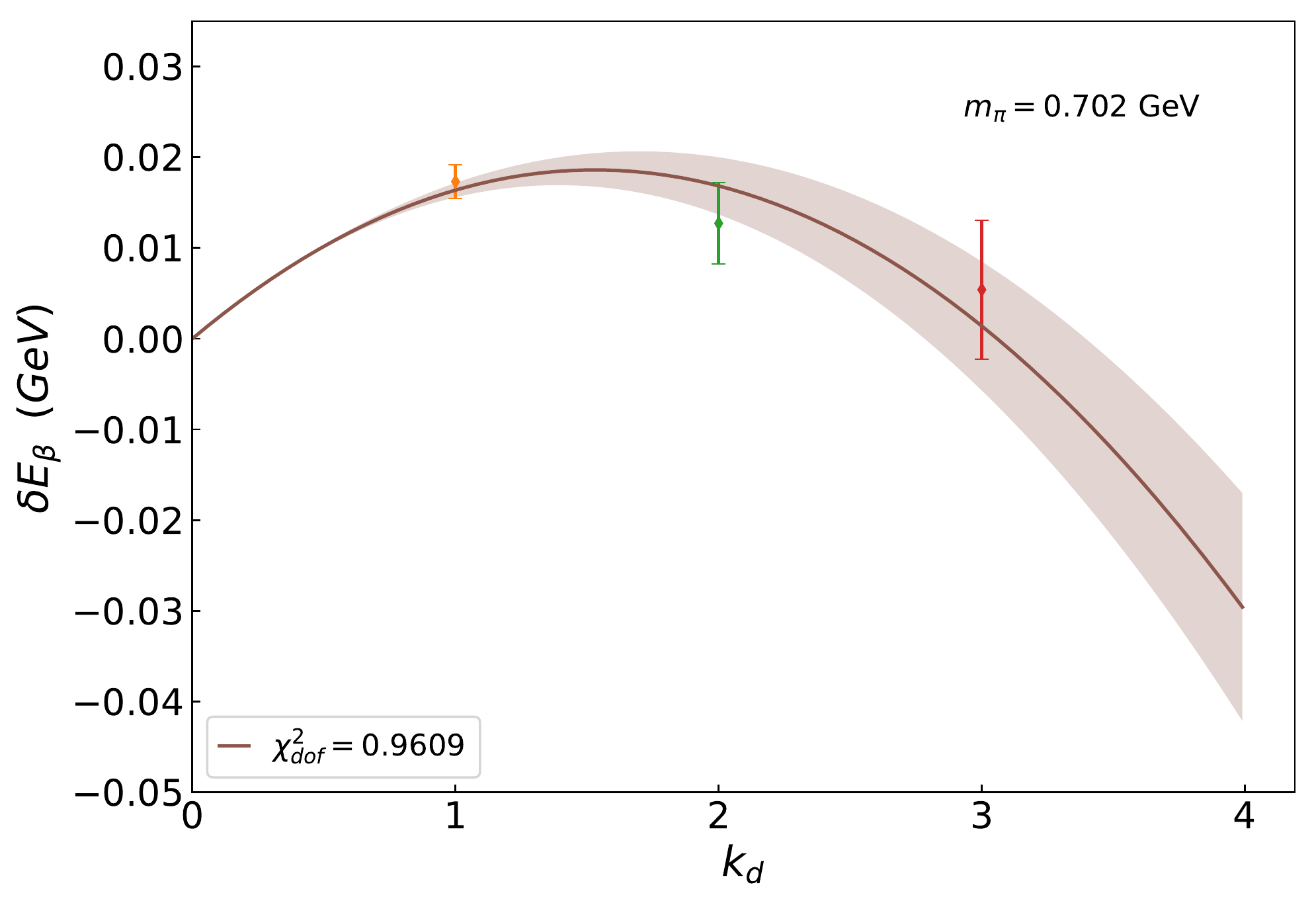}
	\includegraphics[width=0.475\columnwidth]{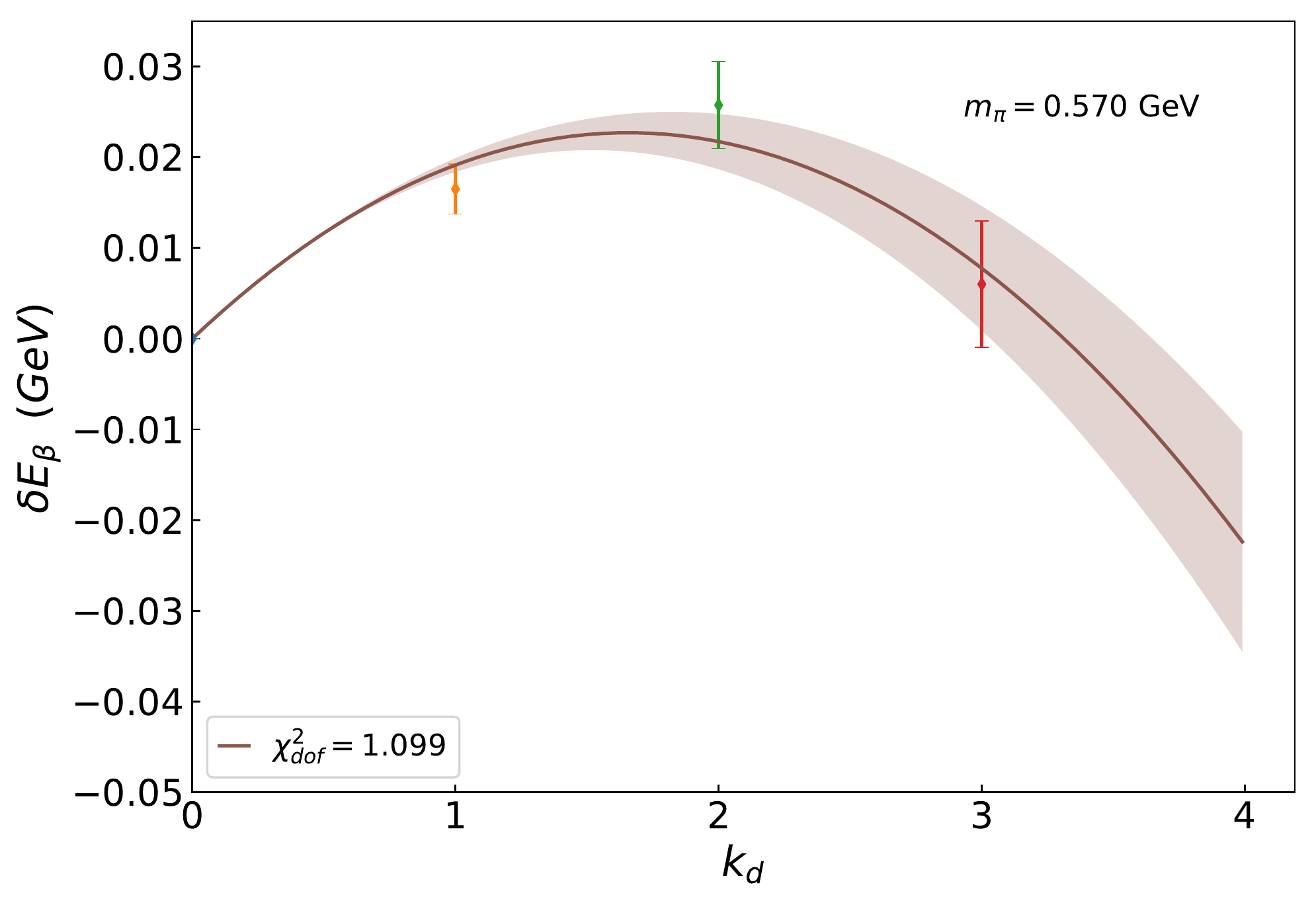}
        \\
        \includegraphics[width=0.475\columnwidth]{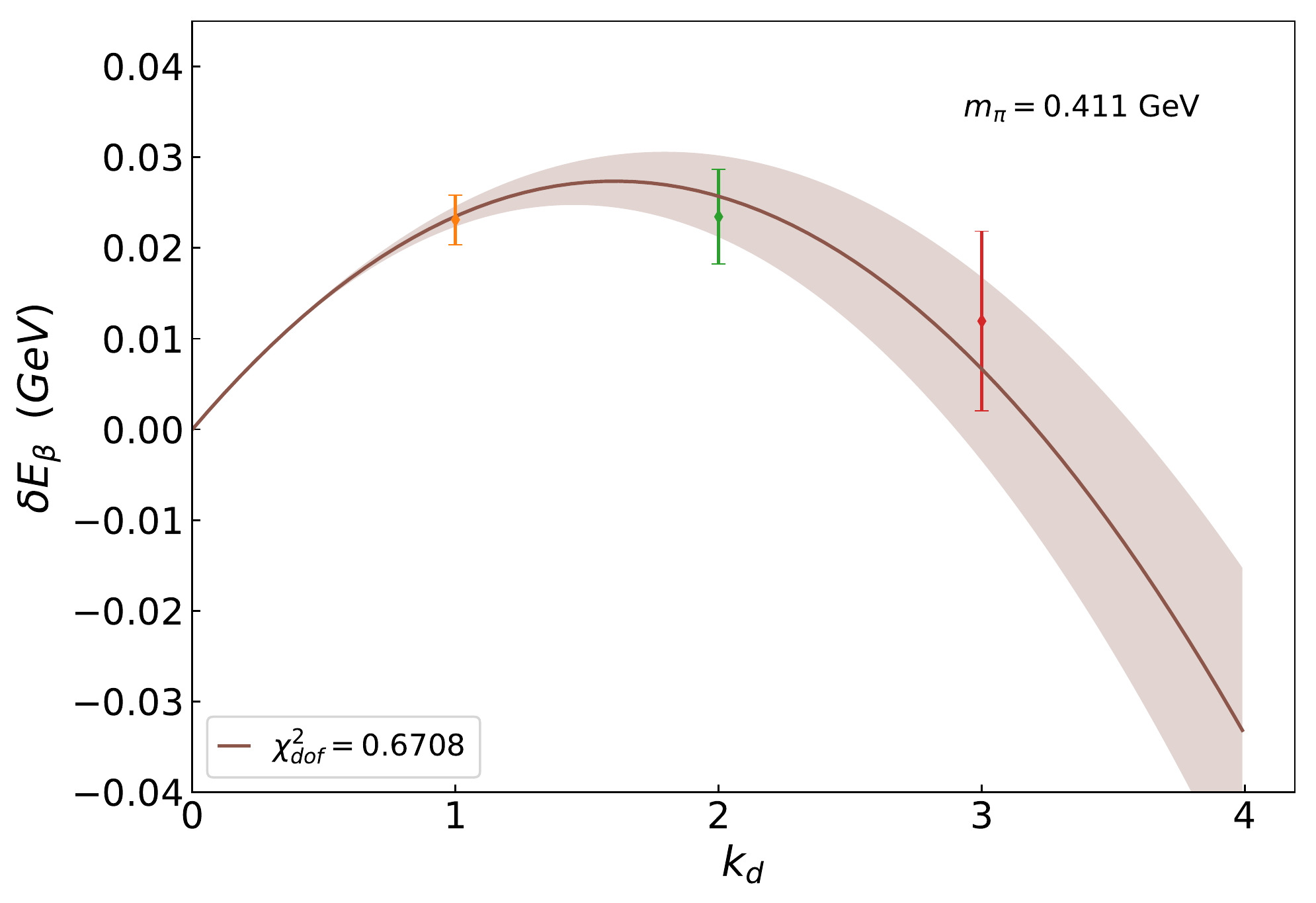}
        \includegraphics[width=0.475\columnwidth]{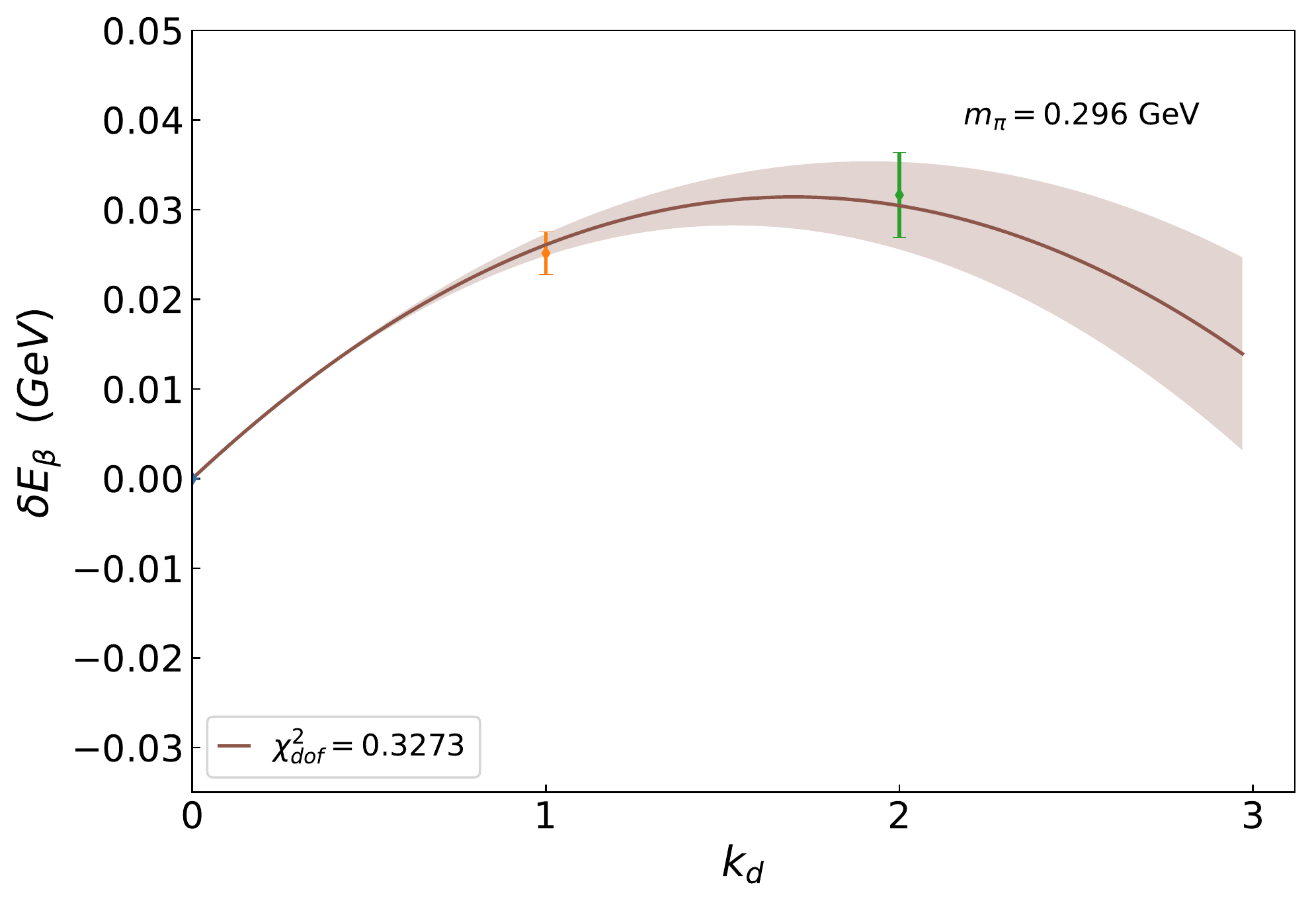}
        \caption{Constrained quadratic fits of the energy shift to the field quanta at each quark mass for the proton.} 
	\label{fig:dEfits4}
\end{figure}
\par
At each pion mass considered, the linearly constrained quadratic fits of \eqnr{eqn:EmmCon} are determined. The final fits are presented in \Fig{fig:dEfits4}. In every case, the full covariance matrix based $\ctdof$ indicates an acceptable fit with the charge of the proton constrained to one.
\par
We have also assumed higher-order terms of the expansion of \eqnr{eqn:EofB} are negligible. In order to check the validity of this assumption, a linearly constrained quadratic $+$ quartic fit incorporating a $c_4\,\kd^4$ term is performed. We find this quartic term provides no additional information, and similar magnetic polarisabilities are observed. When the unconstrained linear $+$ quadratic fit is considered; the linear coefficient $c_1$ produces a charge value $q$ in agreement with one.
\begin{table}[t!]
  \begin{ruledtabular}
    \caption{Magnetic polarisability values for the neutron and proton at each quark mass considered. The number of sources considered for each quark mass varies as described in \Tab{tab:simdet}. The numbers in parentheses describe statistical uncertainties.}
    \label{tab:ResTab}
    \begin{tabular}{ccccc}
      $\mpi$ (GeV) & $\beta^n\left(\text{fm}^3 \times 10^{-4}\right)$ & \multicolumn{1}{l}{${}^n\ctdof$} & $ \beta^p\left( \text{fm}^3 \times 10^{-4}\right)$ & ${}^p\ctdof$ \\ \hline \noalign{\thinspace}
      0.702        & 1.91(12)                                         & 0.85                             & 1.90(19)                                           & 0.96         \\
      0.570        & 1.66(10)                                         & 0.88                             & 1.87(18)                                           & 1.10         \\
      0.411        & 1.53(29)                                         & 0.74                             & 1.98(21)                                           & 0.67         \\
      0.296        & 1.27(37)                                         & 0.91                             & 1.93(22)                                           & 0.33
      \end{tabular}
  \end{ruledtabular}
\end{table}
\par
This is the first time that Euclidean-time plateau fits have been successfully constructed for the proton's magnetic polarisability energy shift. Thus the \SUto quark level eigenmode projection technique is effective in isolating the energy shifts required to access the magnetic polarisability of the proton to be well determined.
\begin{figure}
  \includegraphics[width=\columnwidth]{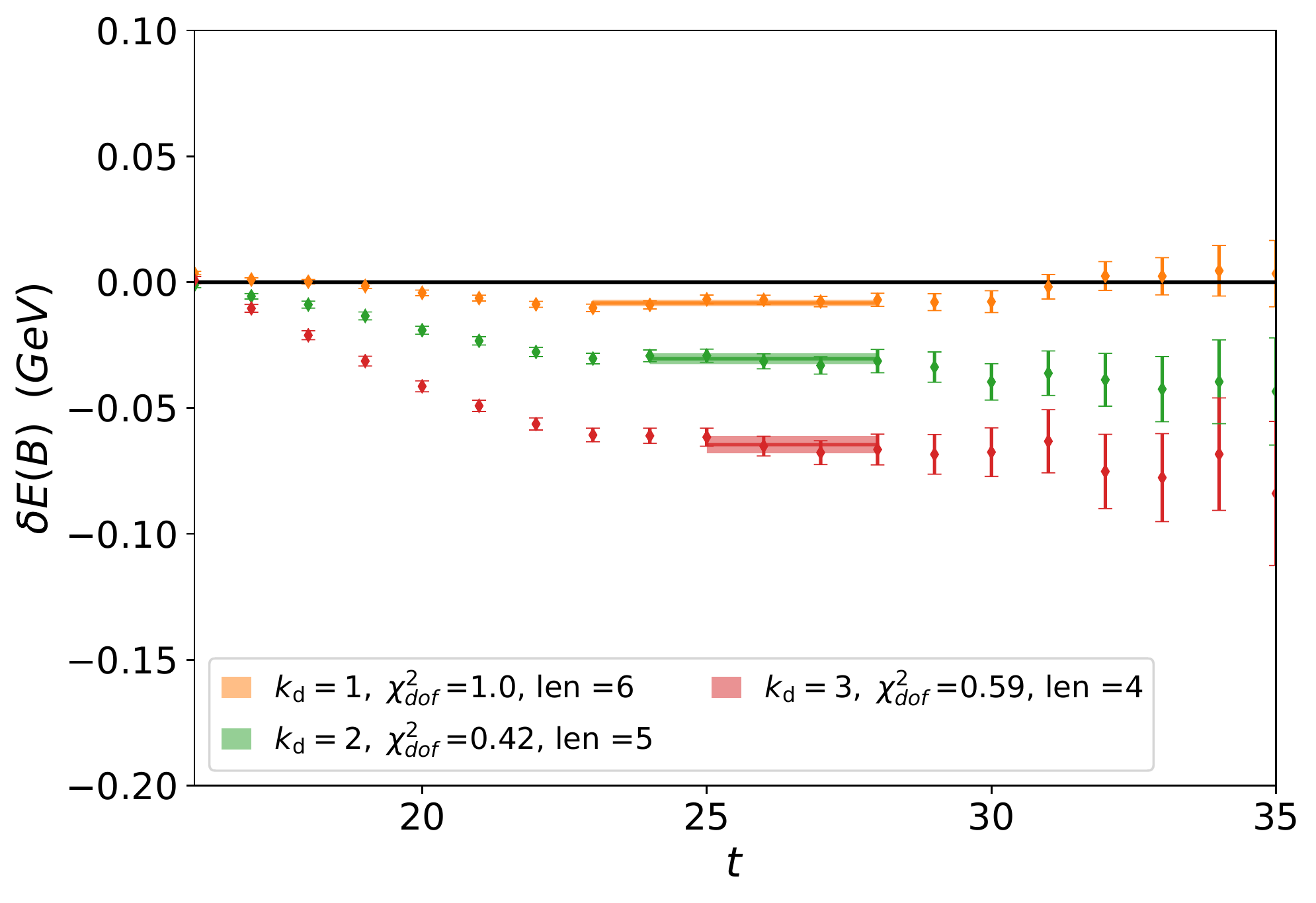}%
  \caption{\label{fig:k2dEpol-n}The magnetic polarisability effective energy shift, $\delta E\rb{B,t}$ of \eqnr{eqn:logR} for the $\mpi=0.570$ GeV neutron as a function of Euclidean time (in lattice units), using a smeared source and the \SUto quark-eigenmode projection technique. Results for field strengths $\kd=1,2,3$ are shown. The selected fits for this ensemble and the $\chi^2_{dof}$ are also illustrated.}
\end{figure}
\begin{figure}
  \includegraphics[width=\columnwidth]{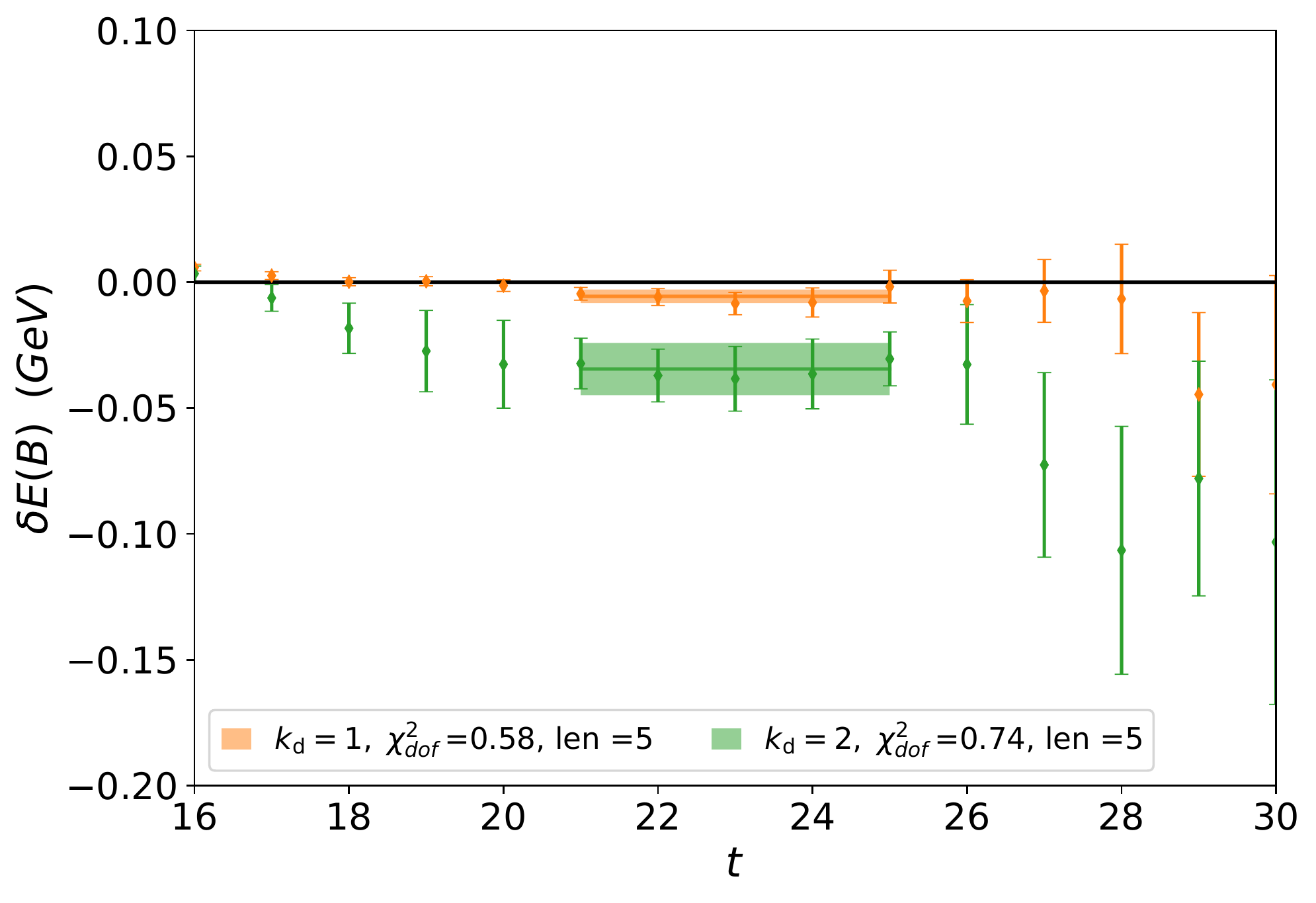}%
  \caption{\label{fig:k4dEpol-n}The magnetic polarisability effective energy shift, $\delta E\rb{B,t}$ of \eqnr{eqn:logR} for the $\mpi=0.296$ GeV neutron as a function of Euclidean time (in lattice units), using a smeared source and the \SUto quark-eigenmode projection technique. Results for field strengths $\kd=1$ and $2$ are shown. The selected fits for this ensemble and the $\chi^2_{dof}$ are also illustrated.}
\end{figure}
\begin{figure}[t]
        \centering
	\includegraphics[width=0.475\columnwidth]{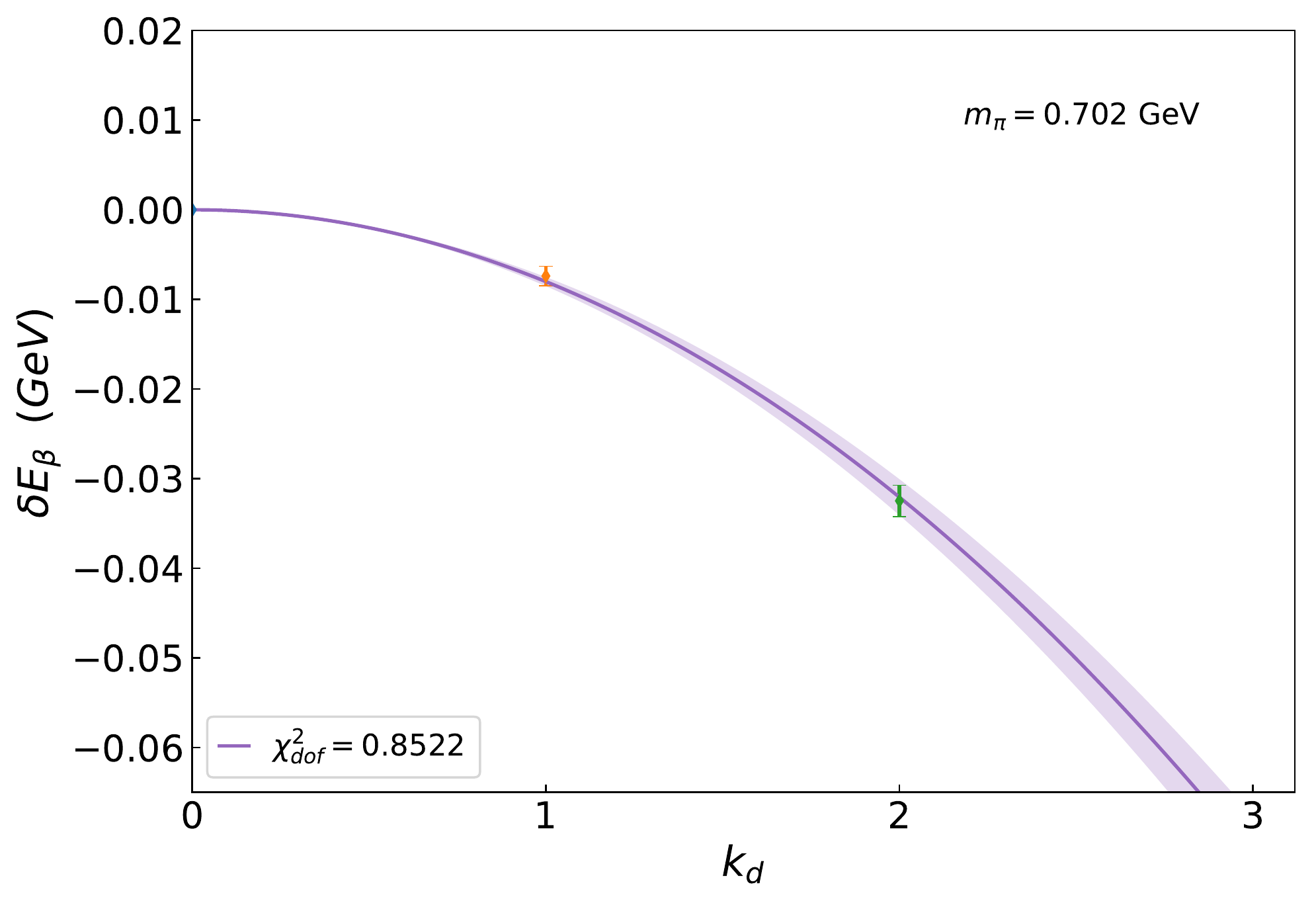}
	\includegraphics[width=0.475\columnwidth]{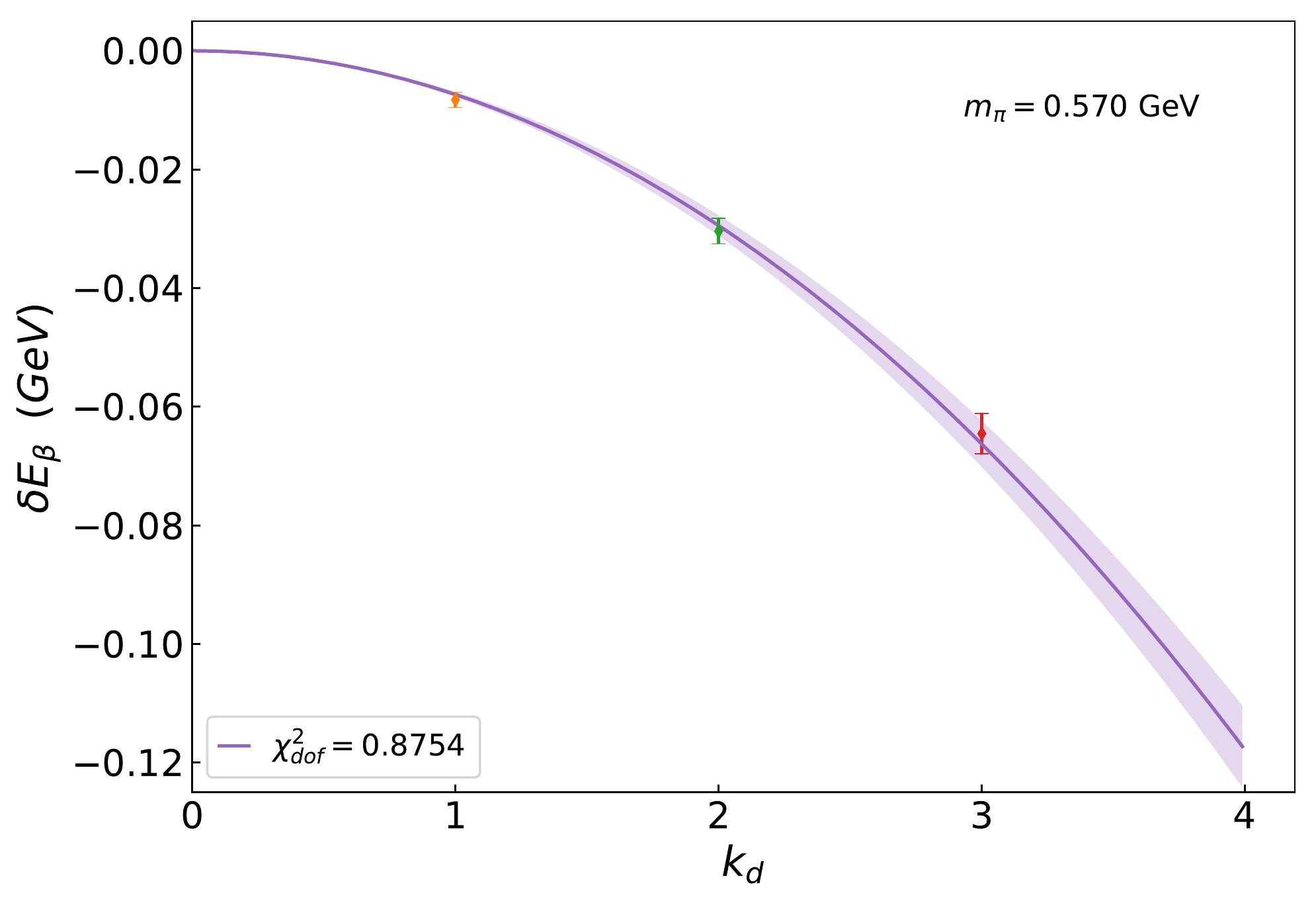}
        \\
        \includegraphics[width=0.475\columnwidth]{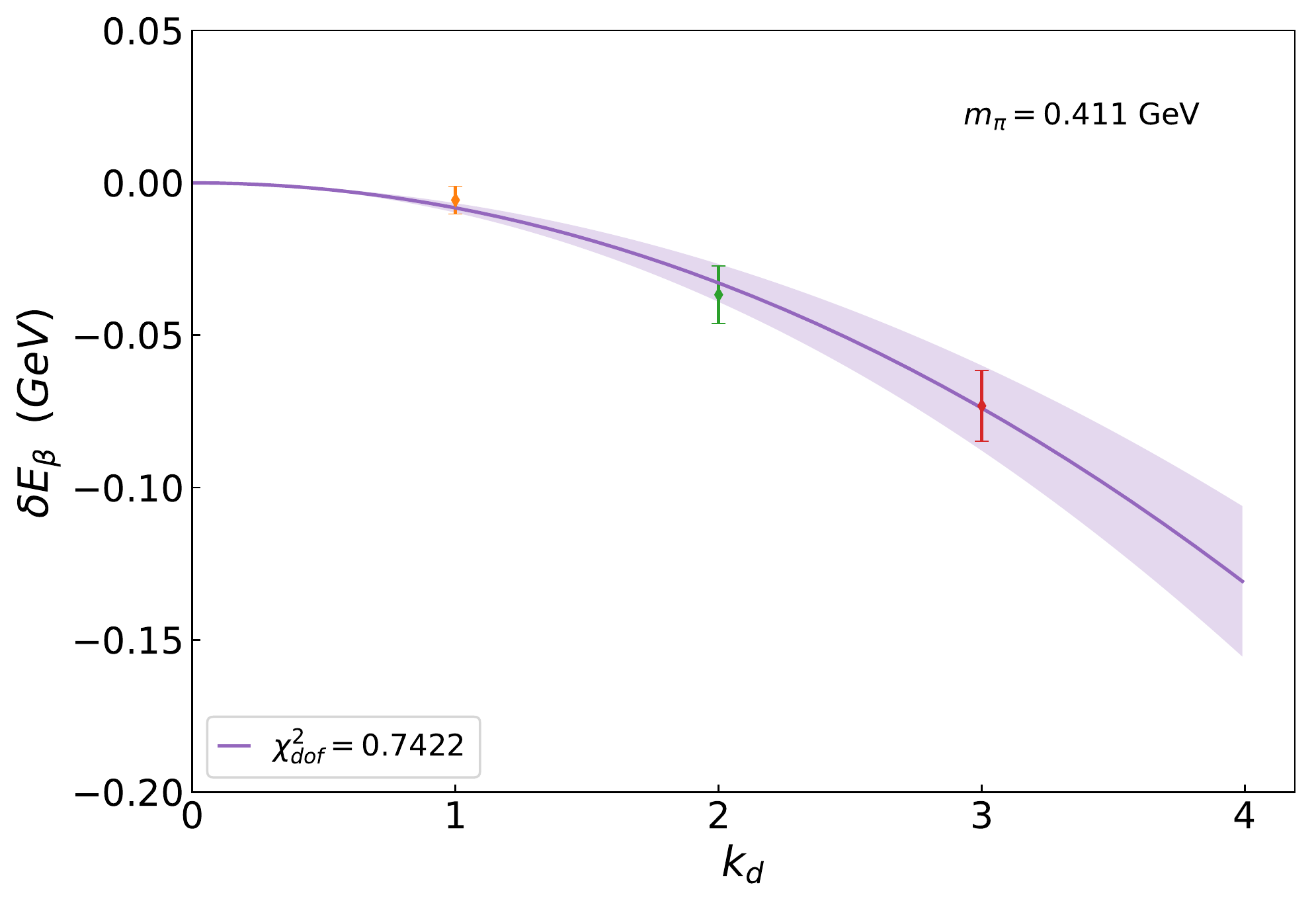}
        \includegraphics[width=0.475\columnwidth]{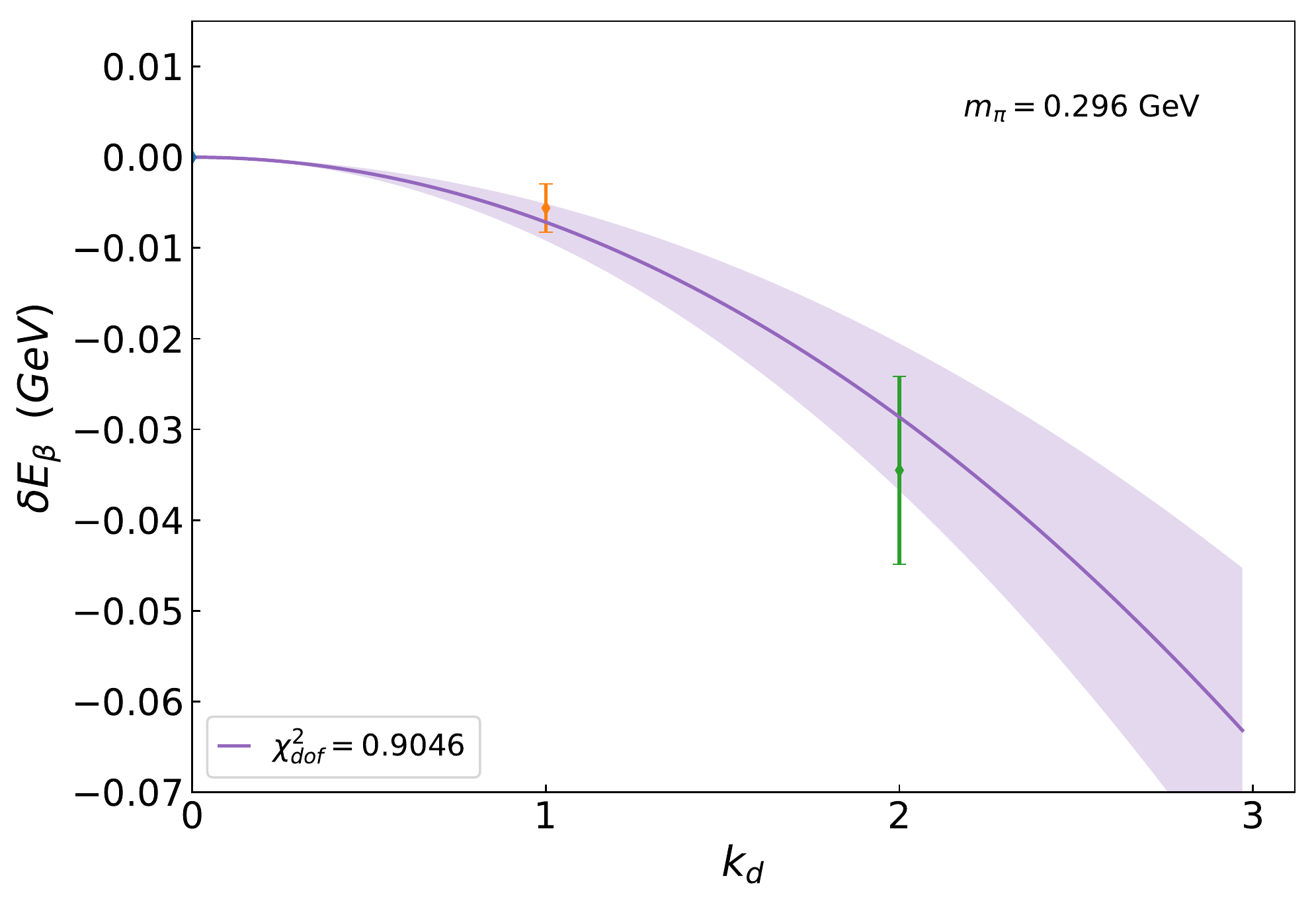}
        \caption{Quadratic fits of the energy shift to the field quanta at each quark mass for the neutron.} 
	\label{fig:dEfits4-n}
\end{figure}
\par
A similar method is followed for the neutron. This time a standard Fourier transform at the hadron sink projects to zero momentum, and as the neutron is overall charge less the subtraction process of \eqnr{eqn:EmmCon} is not required. \Figtwo{fig:k2dEpol-n}{fig:k4dEpol-n} display representative Euclidean-time fits to the effective energy shifts of the neutron, analogous to \Figtwo{fig:k2dEpol}{fig:k4dEpol} for the proton.
\par
The final quadratic fits for the neutron are displayed in \Fig{fig:dEfits4-n}. From the quadratic term of the fit in \eqnr{eqn:EmmCon}, the magnetic polarisability can be found using \eqnr{eqn:betaConv}. Polarisability results for both the proton and neutron are summarised in \Tab{tab:ResTab}.
\par
For the neutron field-strength dependent fit on the $\mpi = 0.702$ GeV ensemble, only the first two non-zero field strengths are used. Fitting to the third requires the additional quartic term and produces a magnetic polarisability value that agrees with a single quadratic fit to the first two field strengths. The neutron polarisability energy shift at the largest field strength considered suffers from the same signal-to-noise problem as that of the proton and is hence not used to extract the magnetic polarisability.
\par
The neutron magnetic polarisabilities obtained herein are in good agreement with those obtained in \Refl{Bignell:2018acn} on the same ensembles. There, a $U(1)$-based Landau-mode projection technique was applied to Landau-gauge fixed quark propagators. We find the \SUto eigenmode quark projection technique to be similarly successful in isolating the neutron ground state in a background magnetic field. Now, for the first time, a unified method for extracting both proton and neutron magnetic polarisabilities has been presented. It is anticipated that the mesons and hyperons will also be tractable using this approach.

\section{Chiral Extrapolation}
\label{sec:chiEFT}
\begin{figure}
  \includegraphics[width=\columnwidth]{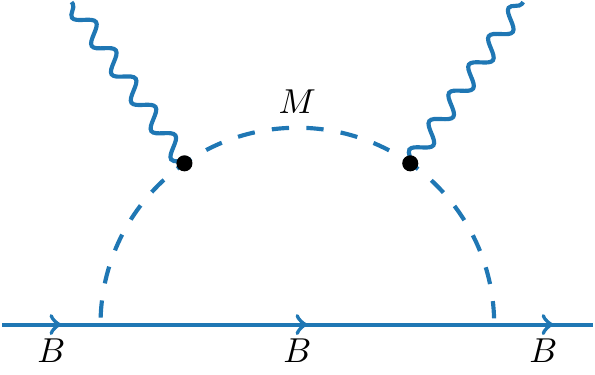}%
  \caption{\label{fig:chiEFT:chiN}The leading-order meson loop contribution to the magnetic polarisability of the nucleon.}
\end{figure}
\begin{figure}
  \includegraphics[width=\columnwidth]{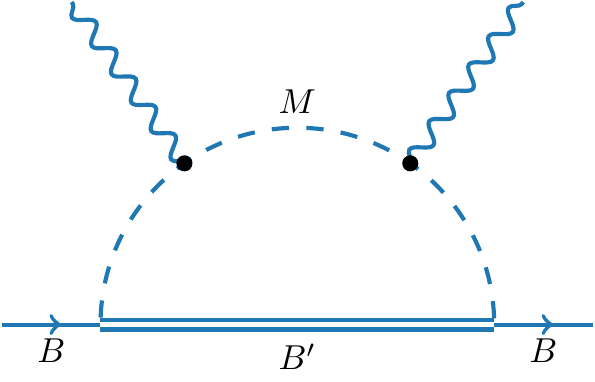}%
  \caption{\label{fig:chiEFT:chiD}The next to leading-order meson loop contribution to the magnetic polarisability of the nucleon, allowing transitions to nearby strongly coupled baryons.}
\end{figure}
To connect lattice results to the physical regime, chiral effective-field theory $\rb{\chi EFT}$ provides a powerful tool. This analysis is a generalisation of \Refl{Hall:2013dva} with modifications arising from the consideration of both the proton and neutron.
\par
The chiral expansion considered is

\begin{align}
  \beta^{B}\rb{\mpit} = &\sum_{M}\,\beta^{M\,B}\rb{\mpit,\Lambda} + \azL + \atL\,\mpit \nonumber \\+ &\sum_{M,\,\Bpr}\,\beta^{M\,\Bpr}\rb{\mpit,\Lambda} + \order{\mpi^3},
  \label{eqn:chiEFT:fitfunc}
\end{align}
where $\azL$ and $\atL$ are residual series coefficients~\cite{Young:2002cj} constrained by our infinite volume corrected lattice QCD results and $\Lambda$ is a renormalisation scale. The leading-order loop contributions $\bpN$ and $\bpD$ are shown in \Figtwo{fig:chiEFT:chiN}{fig:chiEFT:chiD}. \Fig{fig:chiEFT:chiD} allows for transitions of the baryon $B$ to nearby strongly coupled baryons, $\Bpr$, with mass splitting $\Delta$, via a meson, $M$, loop whereas \Fig{fig:chiEFT:chiN} does not encounter any baryon mass-splitting effects.
\par
In the heavy-baryon approximation~\cite{Jenkins:1990jv} appropriate for a low energy expansion, these have integral forms~\cite{Hall:2013dva}
\begin{align}
  \bpN\brmpit = \frac{e^2}{4\,\pi}\frac{1}{288\,\pi^3\,\fpi^2}\chi_{M\,B}\int\dthk\,\frac{\vk^2\,u^2\!\left(k,\Lambda\right)}{\left(\vk^2 + m_{M}^2\right)^3},
  \label{eqn:bpN}
\end{align}
and for $\Delta = m_{\Bpr} -m_B \neq 0$
\begin{multline}
  \bpD\brmpit = \frac{e^2}{4\,\pi}\frac{1}{288\,\pi^3\,\fpi^2}\chi_{M\,\Bpr}\int\dthk\,u^2\!\left(k,\Lambda\right) \nonumber \\
  \times \frac{ \wvk^2\,\Delta\,\left(3\,\wvk + \Delta\right) + \vk^2\,\left(8\,\wvk^2 + 9\,\wvk\,\Delta + 3\,\Delta^2\right)}{8\,\wvk^5\,\left(\wvk+\Delta\right)^3},
\label{eqn:bpD}
\end{multline}
respectively.
Here $\wvk = \sqrt{ \vk^2 + m_{M}^2}$ is the energy carried by the meson $M$ which has three-momentum $\vk$, $\fpi = 92.4$ MeV is the pion decay constant and $u\rb{k,\Lambda}$ is a dipole regulator
\begin{align}
  u\left(k,\Lambda\right) = \frac{1}{\left(1 + \vk^2/\Lambda^2\right)^2},
\end{align}
which ensures that only soft momenta flow through the effective-field theory degrees of freedom.
\par
The renormalised low-energy coefficients of the chiral expansion are formed from the residual series coefficients $\azL$, $\atL$ and the analytic contributions of the loop integrals~\cite{Young:2002ib} which also depend on $\Lambda$. The full details of the renormalisation procedure are provided in the Appendix of \Refl{Young:2002ib}. The standard coefficients for full QCD, $\chi_{M\,B}$ and $\chi_{M\,\Bpr}$ reflect photon couplings to the intermediate meson.
\par
The loop integral of \eqnr{eqn:bpN} for $\bpN$ contains the leading nonanalytic contribution proportional to $1/m_M$. For finite $\Bpr- B$ mass splitting, $\Delta = m_{\Bpr} - m_{B}$, the loop integral of \eqnr{eqn:bpD} accounts for transitions to nearby strongly coupled baryons $\Bpr$ and contributes a nonanalytic logarithmic contribution proportional to $\left( -1/\Delta\right)\,\log{\left(m_{M}/\Lambda\right)}$, to the chiral expansion.
\par
Here we consider mesons $M=\pi$ and $\eta$ for nucleon transitions with $B=n$ or $p$ for the integral of \eqnr{eqn:bpN}. While the total charge of these mesons is zero, it is important to consider their contributions in assessing the contribution of sea-quark-loops. The $\eta^\prime$-meson is much heavier and thus its contribution is suppressed and safely neglected.
\par
We consider transitions to the baryons $\Bpr = \Sigma,\,\Lambda,\,\Delta$ and $\Sigma^{*}$ with mesons $M=\pi,\,\eta$ and $K$. These transitions are accounted for by \eqnr{eqn:bpN} with the appropriate mass splittings and couplings
\par
Our lattice QCD results are electroquenched - they do not include contributions of photon couplings to disconnected sea-quark loops of the vacuum. Disconnected sea-quark loops form part of the full meson dressing of $\chi EFT$  and thus it is necessary to model the corrections associated with their absence in the lattice QCD calculations. Hence the standard coefficients for full QCD $\chi_{M\,B}$ and $\chi_{M\,\Bpr}$ are altered to account for partial quenching effects~\cite{Detmold:2006vu} as explained in \Refl{Hall:2013dva} for the neutron. The proton is briefly discussed below while the neutron follows the analysis in \Refl{Hall:2013dva,Bignell:2018acn}.
\subsection{Partially quenched $\chi$EFT}
\begin{figure*}
  \centering
  \subfloat[The down quark loop diagram where the two photons can couple to valence-valence, sea-sea or valence-sea quarks.\label{fig:pQuenchPiz14}]{
    \includegraphics[width=0.48\textwidth]{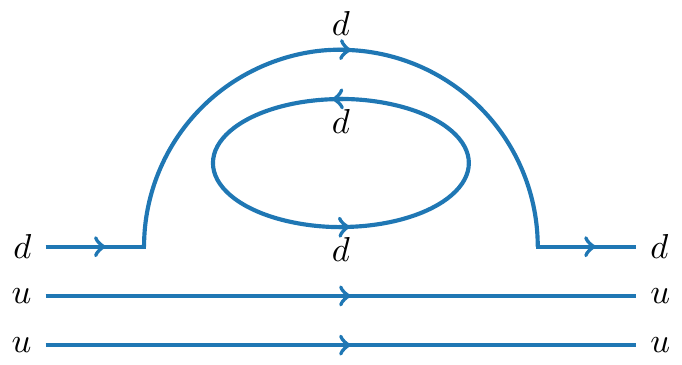}
  }
  \hfill
  \subfloat[The up quark loop diagram where the two photons can couple to valence-valence, sea-sea or valence-sea quarks.\label{fig:pQuenchPiz15}]{
    \includegraphics[width=0.48\textwidth]{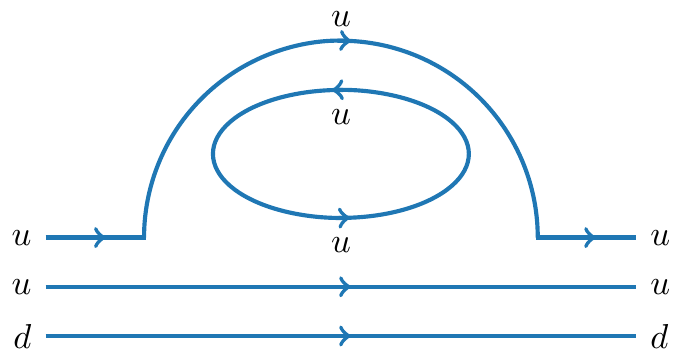}
  }
  \\
  \subfloat[The quark-flow diagram where the two photons can couple only to valence quarks.\label{fig:pQuenchPiz16}]{
    \includegraphics[width=0.48\textwidth]{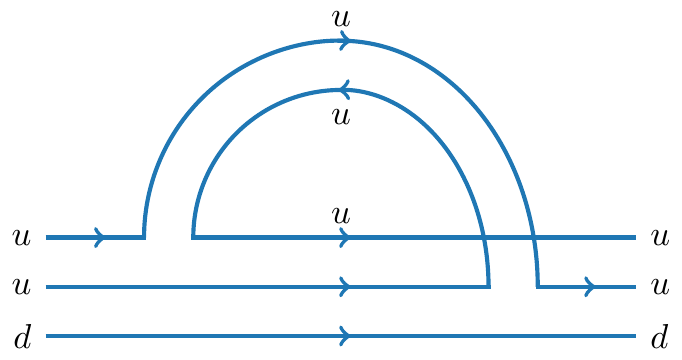}
    }
  \caption{Decomposition of the process $p\rightarrow p\,\piz$ into its possible one-loop quark-flow diagrams. The configuration of the two photon couplings to the valence and/or sea quarks determines the coefficients of partially quenched chiral perturbation theory.}
  \label{fig:chiEFT:QuarkFlow}
\end{figure*}

In order to model the corrections to account for partial quenching effects, the contribution of each quark-flow diagram is separated into 'valence-valence', 'valence-sea and 'sea-sea' contributions. Each of these describes the coupling of the two photons to the valence or sea quarks available in the intermediate state mesons. All possible quark-flow diagrams for the $p\rightarrow p\,\pi^0$ channel are constructed in \Fig{fig:chiEFT:QuarkFlow} without attaching external photons to the meson. As there is baryon no mass splitting, this is an example of \Fig{fig:chiEFT:chiN}.
\par
As \Figtwo{fig:pQuenchPiz14}{fig:pQuenchPiz15} have both sea and valence quark lines of the intermediate meson, photon lines may be attached to the valence or sea-quark lines of the intermediate meson. Hence they may contribute to all three sectors. This is in contrast to \Fig{fig:pQuenchPiz16} which contains only valence quarks and hence contributes only to the valence-valence sector. The contributions to each sector is proportional to the quark charges, i.e. for \Fig{fig:pQuenchPiz15} the leading non-analytic term of the chiral expansion has coefficents
\begin{align}
  \chi_{v-v} &\propto \qu^2, \\
  \chi_{v-s} &\propto 2\,\qu\,\qub \label{eqn:factwo}\\
  \chi_{s-s} &\propto \qub^2, 
\end{align}
where \eqnr{eqn:factwo} reflects the two orderings of the photon couplings available.

\par
While the sum of the valence-valence, valence-sea and sea-sea contributions is zero for this process due to the neutrality of the $\pi^0$ meson; the valence-sea and sea-sea terms are not present in the lattice QCD simulation and hence must be accounted for.
\par
The sea-sea disconnected sea-quark-loop flow for Diagram \ref{fig:pQuenchPiz14} can be isolated by temporarily replacing the down-quark loop with a strange quark~\cite{Leinweber:2002qb}. This provides a coupling strength
\begin{align}
  \chi_{s-s}^{diag(b)} \propto \,q^2_{\overline{d}}\,\chi^2_{K^0\Sigma^+} = q^2_{\overline{d}}\,\,2\,\left(D-F\right)^2.
\end{align}
Repeating the above procedure for the up-quark loop of Diagram \ref{fig:pQuenchPiz15} one finds
\begin{align}
  \chi_{s-s}^{diag(c)} \propto& \,q_{\overline{u}}^2\,\rb{\chi^2_{K^+\Lambda} + \chi^2_{\Sigma^0 K^+}} \nonumber \\=& q_{\overline{u}}^2\,\rb{ \frac{1}{3}\,\left(3\,F+D\right)^2 + \left(D-F\right)^2 }.
\end{align}
\par
The components of the $p\rightarrow p\,\pi^0$ channel which have a disconnected sea-quark loop have been identified and hence the sea-sea contributions have been calculated. The same process may be performed for the valence-sea contributions. As the total contribution is known from standard $\chi$PT, the remaining valence-valence contribution which includes the connected quark-flow diagram of Diagram \ref{fig:chiEFT:QuarkFlow}(a) is also known. All such channels for the integral processes described by \eqnr{eqn:chiEFT:fitfunc} are investigated using the diagrammatic procedure described above for $p\rightarrow p\,\pi^0$.

\begin{table*}[t]
  \caption{Chiral coefficients for the leading-order loop integral contributions for the proton.}
  \label{tab:App:Lat:pChiEFT}
  \makebox[1\textwidth][c]{
    \resizebox{1.0\textwidth}{!}{
      \begin{tabular}{llll}
        \hline\hline\noalign{\vskip 2mm}
    Process    & Total    & Valence-sea    & Sea-sea    \\ \hline
    $p \rightarrow N\,\pi$    &     &     &     \\ \hline\noalign{\vskip 2mm}
    $p \rightarrow p\,\piz$    & $0$    & $\frac{3}{6} \, \rb{ 2\,\qu\,\qub\,\rb{ \chiKpSigz + \chiKpLambda } + 2\,\qd\,\qdb\,\chiKzSigmap }$    & $\frac{3}{6} \, \rb{ \qub^2\,\rb{ \chiKpSigz + \chiKpLambda } + \qdb^2\,\chiKzSigmap }$    \\
    $p \rightarrow p\,\etaCust$    & $0$    & $\frac{1}{6} \, \rb{ 2\,\qu\,\qub\,\rb{ \chiKpSigz + \chiKpLambda } + 2\,\qd\,\qdb\,\chiKzSigmap }$    & $\frac{1}{6} \, \rb{ \qub^2\,\rb{ \chiKpSigz + \chiKpLambda } + \qdb^2\,\chiKzSigmap }$    \\
    $p \rightarrow p\,\etaCust^\prime$    & $0$    & $\frac{2}{6} \, \rb{ 2\,\qu\,\qub\,\rb{ \chiKpSigz + \chiKpLambda } + 2\,\qd\,\qdb\,\chiKzSigmap }$    & $\frac{2}{6} \, \rb{ \qub^2\,\rb{ \chiKpSigz + \chiKpLambda } + \qdb^2\,\chiKzSigmap }$    \\
    $p \rightarrow n\,\pip$    & $\chinpip$    & $2\,\qu\,\qdb\,\rb{ \chiKpSigz + \chiKpLambda }$    & $\qdb^2\,\rb{ \chiKpSigz + \chiKpLambda }$    \\
    $p \rightarrow p^{+}\,\pi^{-}$    & $0$    & $2\,\qd\,\qub\,\chiKzSigmap$    & $\qub^2\,\chiKzSigmap$    \\ \noalign{\vskip 2mm}\hline
    $p \rightarrow \Sigma\,K$    &     &     &     \\ \hline\noalign{\vskip 2mm}
    $p \rightarrow \rb{\Sigz, \Lambda}\,\Kp$    & $\chiKpSigz + \chiKpLambda$    & $2\,\qu\,\qsb\,\rb{ \chiKpSigz + \chiKpLambda }$    & $\qsb^2\,\rb{ \chiKpSigz + \chiKpLambda }$    \\
    $p \rightarrow \Sigmap\,\Kz$    & $0$    & $2\,\qd\,\qsb\,\chiKzSigmap$    & $\qsb^2\,\chiKzSigmap$    \\ \noalign{\vskip 2mm}\hline
    $p \rightarrow \Delta\,\pi$    &     &     &     \\ \hline\noalign{\vskip 2mm}
    $p \rightarrow \Deltaz\,\pip$    & $\chiPipDeltaz$    & $2\,\qu\,\qdb\,\chiKpSigzStar$    & $\qdb^2\,\chiKpSigzStar$    \\
    $p \rightarrow \Deltapp\,\pim$    & $\chiPimDeltapp$    & $2\,\qd\,\qub\,\chiKzSigmapStar$    & $\qub^2\,\chiKzSigmapStar$    \\
    $p \rightarrow \Deltap\,\piz$    & $0$    & $\frac{3}{6}\,\rb{2\,\qu\,\qub\,\chiKpSigzStar + 2\,\qd\,\qdb\,\chiKzSigmapStar}$    & $\frac{3}{6}\, \rb{ \qub^2\,\chiKpSigzStar + \qdb^2\,\chiKzSigmapStar }$    \\
    $p \rightarrow \Deltap\,\etaCust$    & $0$    & $\frac{1}{6}\,\rb{2\,\qu\,\qub\,\chiKpSigzStar + 2\,\qd\,\qdb\,\chiKzSigmapStar}$    & $\frac{1}{6}\, \rb{ \qub^2\,\chiKpSigzStar + \qdb^2\,\chiKzSigmapStar }$    \\
    $p \rightarrow \Deltap\,\etaCust^\prime$    & $0$    & $\frac{2}{6}\,\rb{2\,\qu\,\qub\,\chiKpSigzStar + 2\,\qd\,\qdb\,\chiKzSigmapStar}$    & $\frac{2}{6}\, \rb{ \qub^2\,\chiKpSigzStar + \qdb^2\,\chiKzSigmapStar }$    \\ \noalign{\vskip 2mm}\hline
    $p \rightarrow \Sigma^{*}\,K$    &     &     &     \\ \hline\noalign{\vskip 2mm}
    $p \rightarrow \SigzStar\,\Kp$    & $\chiKpSigzStar$    & $2\,\qu\,\qsb\,\chiKpSigzStar$    & $\qsb^2\,\chiKpSigzStar$    \\
    $p \rightarrow \SigmapStar\,\Kz$    & $0$    & $2\,\qd\,\qsb\,\chiKzSigmapStar$    & $\qsb^2\,\chiKzSigmapStar$    \\    \noalign{\vskip 2mm}\hline\hline
  \end{tabular}
  }
  }
\end{table*}

\begin{table*}[t]
  \caption{Chiral coefficients for the leading-order loop integral contributions for the neutron.}
  \label{tab:App:Lat:nChiEFT}
  \makebox[1\textwidth][c]{
    \resizebox{1.0\textwidth}{!}{
      \begin{tabular}{llll}
        \hline\hline
    Process    & Total    & Valence-sea    & Sea-sea    \\ \hline
    $n \rightarrow N\,\pi$    &     &     &     \\ \hline\noalign{\vskip 2mm}
    $n \rightarrow n\,\piz$    & $0$    & $\frac{3}{6}\,\rb{2\,\qu\,\qub\,\chiKpSigmam + 2\,\qd\,\qdb\,\rb{ \chiKzSigz + \chiKzLambda }}$    & $\frac{3}{6}\,\rb{\qub^2\,\chiKpSigmam + \qdb^2\,\rb{ \chiKzSigz + \chiKzLambda }}$    \\
    $n \rightarrow n\,\etaCust$    & $0$    & $\frac{1}{6}\,\rb{2\,\qu\,\qub\,\chiKpSigmam + 2\,\qd\,\qdb\,\rb{ \chiKzSigz + \chiKzLambda }}$    & $\frac{1}{6}\,\rb{\qub^2\,\chiKpSigmam + \qdb^2\,\rb{ \chiKzSigz + \chiKzLambda }}$    \\
    $n \rightarrow n\,\etaCust^\prime$    & $0$    & $\frac{2}{6}\,\rb{2\,\qu\,\qub\,\chiKpSigmam + 2\,\qd\,\qdb\,\rb{ \chiKzSigz + \chiKzLambda }}$    & $\frac{2}{6}\,\rb{\qub^2\,\chiKpSigmam + \qdb^2\,\rb{ \chiKzSigz + \chiKzLambda }}$    \\
    $n \rightarrow p\,\pi^{-}$    & $\chiPimp$    & $2\,\qd\,\qub\,\rb{ \chiKzSigz + \chiKzLambda }$    & $\qub^2\,\rb{ \chiKzSigz + \chiKzLambda }$    \\
    $n \rightarrow n^{-}\,\pi^{+}$    & $0$    & $2\,\qu\,\qdb\,\chiKpSigmam$    & $\qdb^2\,\chiKpSigmam$    \\ \noalign{\vskip 2mm}\hline
    $n \rightarrow \Sigma\,K$    &     &     &     \\ \hline\noalign{\vskip 2mm}
    $n \rightarrow \rb{\Sigz,\,\Lambda}\,\Kz$    & $0$    & $2\,\qd\,\qsb\,\rb{ \chiKzSigz + \chiKzLambda }$    & $\qsb^2\,\rb{ \chiKzSigz + \chiKzLambda }$    \\
    $n \rightarrow \Sigmam\,\Km$    & $\chiKpSigmam$    & $2\,\qu\,\qsb\,\chiKpSigmam$    & $\qsb^2\,\chiKpSigmam$    \\ \noalign{\vskip 2mm}\hline
    $n \rightarrow \Delta\,\pi$    &     &     &     \\ \hline\noalign{\vskip 2mm}
    $n \rightarrow \Deltaz\,\piz$    & $0$    & $\frac{3}{6}\,\rb{2\,\qu\,\qub\,\chiKpSigmamStar + 2\,\qd\,\qdb\,\chiKzSigzStar}$    & $\frac{3}{6}\,\rb{\qub^2\,\chiKpSigmamStar + \qdb^2\,\chiKzSigzStar}$    \\
    $n \rightarrow \Deltaz\,\etaCust$    & $0$    & $\frac{1}{6}\,\rb{2\,\qu\,\qub\,\chiKpSigmamStar + 2\,\qd\,\qdb\,\chiKzSigzStar}$    & $\frac{1}{6}\,\rb{\qub^2\,\chiKpSigmamStar + \qdb^2\,\chiKzSigzStar}$    \\
    $n \rightarrow \Deltaz\,\etaCust^\prime$    & $0$    & $\frac{2}{6}\,\rb{2\,\qu\,\qub\,\chiKpSigmamStar + 2\,\qd\,\qdb\,\chiKzSigzStar}$    & $\frac{2}{6}\,\rb{\qub^2\,\chiKpSigmamStar + \qdb^2\,\chiKzSigzStar}$    \\
    $n \rightarrow \Deltap\,\pim$    & $\chiDeltappim$    & $2\,\qd\,\qub\,\chiKzSigzStar$    & $\qub^2\,\chiKzSigzStar$    \\
    $n \rightarrow \Deltam\,\pip$    & $\chiDeltampip$    & $2\,\qu\,\qdb\,\chiKpSigmamStar$    & $\qdb^2\,\chiKpSigmamStar$    \\ \noalign{\vskip 2mm}\hline
    $n \rightarrow \Sigma^{*}\,K$    &     &     &     \\ \hline\noalign{\vskip 2mm}
    $n \rightarrow \SigzStar\,\Kz$    & $0$    & $2\,\qd\,\qsb\,\chiKzSigzStar$    & $\qsb^2\,\chiKzSigzStar$    \\
    $n \rightarrow \SigmamStar\,\Kp$    & $\chiKPSigmamStar$    & $2\,\qu\,\qsb\,\chiKpSigmamStar$    & $\qsb^2\,\chiKpSigmamStar$    \\ \noalign{\vskip 2mm}\hline\hline
  \end{tabular}
  }
  }
\end{table*}

\begin{table*}[]
  \centering
  \caption{$SU(3)$ flavour coupling coefficients for the chiral effective field theory analysis. The header row indicates the intermediate baryon species in the meson-baryon loop dressing. Through conservation of quark flavour, one can identify the baryon which is being dressed.}
  \label{tab:App:Lat:chitCoeff}
    \makebox[1\textwidth][c]{
      \resizebox{1.0\textwidth}{!}{
        \begin{tabular}{lllllllllllllllllll}
          \hline\hline\noalign{\vskip 2mm}
          \multicolumn{3}{c}{$\Sigma^{*}$}                     &         & \multicolumn{3}{c}{$\Delta$}                       &         & \multicolumn{3}{c}{$\Sigma$}               &         & \multicolumn{3}{c}{$\Lambda$}                           &         & \multicolumn{3}{c}{$N$}                \\ \hline\noalign{\vskip 2mm}
          $\chiKpSigmamStar$ & $\quad$ & $\frac{4}{9}\,\mcC^2$ & $\quad$ & $\chiDeltappim$  & $\quad$ & $\frac{4}{9}\,\mcC^2$ & $\quad$ & $\chiKpSigmam$ & $\quad$ & $2\,\rb{D-F}^2$ & $\quad$ & $\chiKzLambda$ & $\quad$ & $\frac{1}{3}\,\rb{D+3\,F}^2$ & $\quad$ & $\chinpip$ & $\quad$ & $2\,\rb{D+F}^2$ \\
          $\chiKzSigzStar$   &         & $\frac{2}{9}\,\mcC^2$ &         & $\chiDeltampip$  &         & $\frac{4}{3}\,\mcC^2$ &         & $\chiKzSigz$   &         & $\phantom{2\,}\rb{D-F}^2$    &         & $\chiKpLambda$ &         & $\frac{1}{3}\,\rb{D+3\,F}^2$ &         & $\chiPimp$ &         & $2\,\rb{D+F}^2$ \\
          $\chiKzSigmapStar$ &         & $\frac{4}{9}\,\mcC^2$ &         & $\chiPipDeltaz$  &         & $\frac{4}{9}\,\mcC^2$ &         & $\chiKpSigz$   &         & $\phantom{2\,}\rb{D-F}^2$    &         &                &         &                              &         &            &         &                 \\
          $\chiKpSigzStar$   &         & $\frac{2}{9}\,\mcC^2$ &         & $\chiPimDeltapp$ &         & $\frac{4}{3}\,\mcC^2$ &         & $\chiKzSigmap$ &         & $2\,\rb{D-F}^2$ &         &                &         &                              &         &            &         &\\\noalign{\vskip 2mm}\hline\hline
          \end{tabular}
      }
    }
\end{table*}
\par
Having performed this procedure for each relevant channel, the coefficients used when fitting the lattice QCD results reflect the absence of the disconnected sea-quark-loop contributions and can be determined by subtracting the valence-sea and sea-sea contributions from the total contribution in \Tabtwo{tab:App:Lat:pChiEFT}{tab:App:Lat:nChiEFT} for the proton and neutron respectively. The numerical value of the coefficients can be found in \Tab{tab:App:Lat:chitCoeff} where the standard values of $g_A = 1.267$ and $\mathcal{C} = 1.52$ with $g_A = D + F$ and the SU(6) symmetry relation $F=\frac{2}{3}\,D$ are used.
\par
The regulator mass, $\Lambda = 0.80$ GeV~\cite{Wang:2008vb,Young:2002cj,Leinweber:2004tc,Leinweber:2006ug,Wang:1900ta} is chosen in anticipation of accounting for the missing disconnected sea-quark-loop contributions in the lattice QCD calculations. This regulator mass enables corrections to the pion cloud contributions associated with missing disconnected sea-quark-loop contributions as it defines a pion cloud contribution to masses~\cite{Young:2002cj}, magnetic moments~\cite{Leinweber:2004tc} and charge radii~\cite{Wang:2008vb}. The nucleon core contribution is insensitive to sea-quark-loop contributions at this regulator mass~\cite{Wang:2013cfp}.

\par
To consider the effect of the finite-volume of the lattice, we replace the continuum integrals of the chiral expansion with sums over the momenta available on the periodic lattice. It is important to note that the lattice volume is slightly different on each of the four lattice ensembles used due to our use of the Sommer scale. To produce inite-volume corrected (FVC) results, $\beta^{FVC}_{v-v}$, we take the difference between these sums and the continuum integrals
\begin{align}
  &\beta^{FVC}_{v-v}\rb{\mpit} = \beta^{lat.}_{v-v}\rb{\mpit} \nonumber \\&-\rb{\sum_{M}\,\beta_{SUM}^{M\,B}\rb{\mpit,\LamFV} + \sum_{M,\,\Bpr}\beta_{SUM}^{M\,\Bpr}\rb{\mpit,\LamFV} } \nonumber \\ &+\rb{\sum_{M}\beta^{M\,B}\rb{\mpit,\LamFV} + \sum_{M,\,\Bpr}\,\beta^{M\,\Bpr}\rb{\mpit,\LamFV} }, 
  \label{SU3U1:eqn:betaInfVCorr}
\end{align}
where we note that we are correcting for finite-volume only and hence the coefficients used in evaluating these sums and integrals reflect only valence-valence contributions. The finite-volume corrections should be independent of the value of the regulator parameter, $\LamFV$, as long as $\LamFV$ is sufficiently large. Here we choose $\LamFV = 2.0$ GeV~\cite{Hall:2010ai}.
\par
The strength of the $\chi PT$ analysis is that the leading and next-to-leading non-analytic terms of the chiral expansion are model-independent predictions of chiral perturbation theory. The leading source of uncertainty comes from the higher order terms in the expansion. We provide an estimation of the uncertainty in these terms through variation of the regulator parameter $\Lambda$ over a wide range.

\subsection{Analysis}
The extrapolation to the physical regime requires that the residual series coefficients $\azL$ and $\atL$ are constrained by fitting to the finite-volume corrected lattice results
\begin{align}
  \azL + \atL\,\mpit &= \beta_{v-v}^{FVC}\rb{\mpit} \nonumber - \sum_M\,\beta_{v-v}^{M\,B}\rb{\mpit,\,\Lambda} \nonumber \\&- \sum_{M,\,\Bpr}\,\beta_{v-v}^{M\,\Bpr}\rb{\mpit,\,\Lambda},
  \label{SU3U1:eqn:a0a2fit}
\end{align}
where the regulator parameter $\Lambda$ takes the value $\Lambda=0.80$ GeV as discussed above. After the residual series coefficients have been determined, the chiral expansion of \eqnr{eqn:chiEFT:fitfunc} can be used to calculate the magnetic polarisability for any value for $\mpit$. Valence-sea and sea-sea loop integral contributions are accounted for by using the chiral coefficients for the \enquote{total} process. A physical extrapolation is produced by setting $\mpi = \mpi^{phys} = 0.140$ GeV.
\par
The resulting chiral extrapolation for the proton predicts a magnetic polarisability of $\beta^{p} = 2.79(22) \times 10^{-4}$ fm$^3$ where the numbers in parentheses represents the statistical uncertainty. By considering the variation of the regulator parameter over the broad range $0.6 \text{ GeV}\leq \Lambda \leq 1.0$ GeV a systematic uncertainty associated with the higher order terms of the chiral expansion can be reported. Thus the prediction for the magnetic polarisability of the proton at the physical point is
\begin{align*}
  \beta^{p} = 2.79(22)\rb{{}^{+13}_{-18}} \times 10^{-4} \text{ fm}^3.
\end{align*}
\begin{figure}
  \includegraphics[width=\columnwidth]{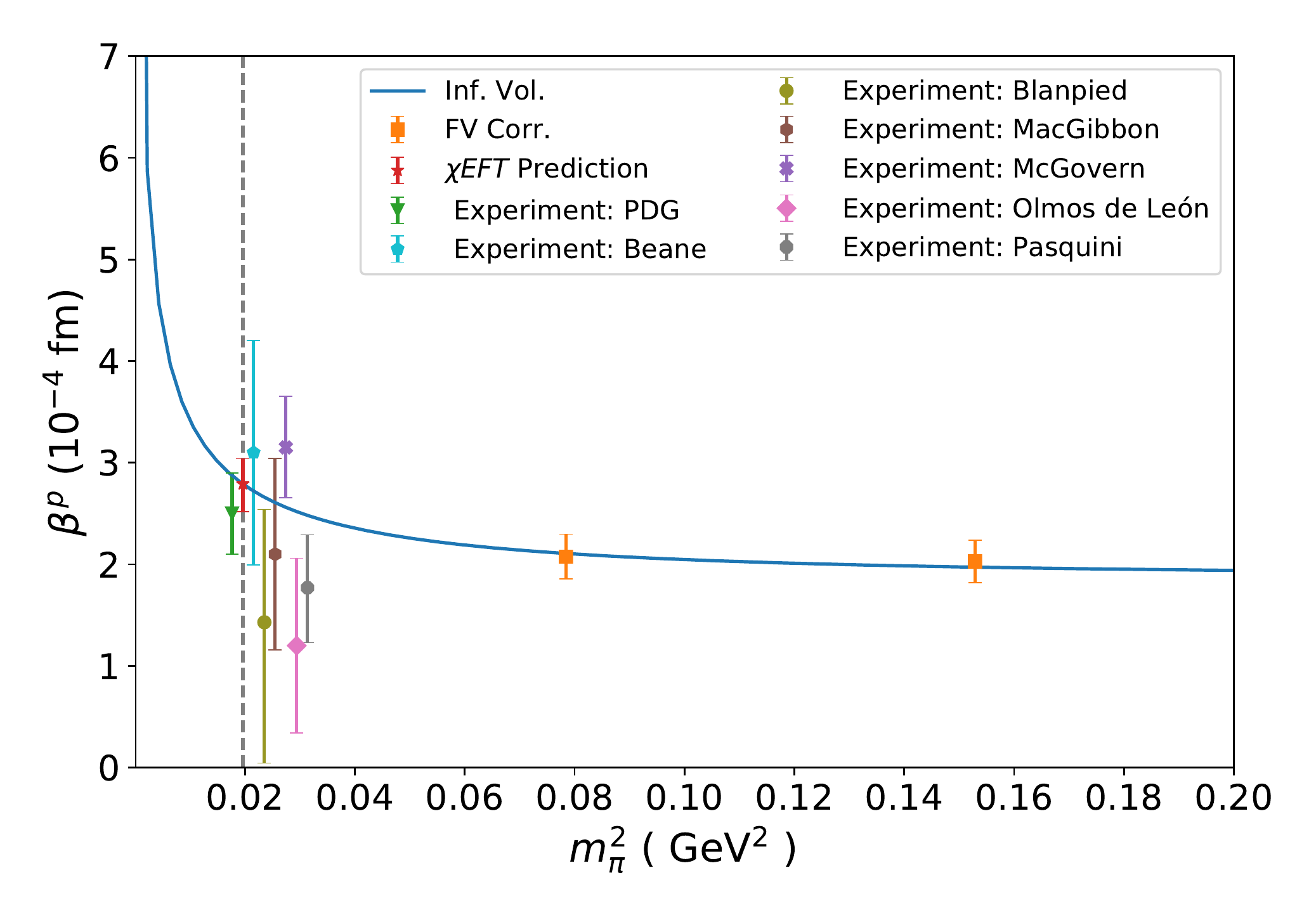}%
  \caption{\label{fig:chiEFT:compP}The magnetic polarisability of the proton, $\beta^p$, from the chiral effective field analysis herein $\rb{\chi EFT\text{ Prediction}}$. Lattice results of this work, finite-volume corrected (FV Corr.) with total QCD coefficients are compared with experimental measurements via an infinite-volume (Inf. Vol) chiral extrapolation with total QCD coefficents. The error bar at the physical point reflects systematic and statistical uncertainties added in quadrature. Experimental results from the PDG~\cite{Tanabashi:2018oca}, McGovern \emph{et~al.}\cite{McGovern:2012ew}, Beane \emph{et~al.}\cite{Beane:2002wn}, Blanpied \emph{et~al.}\cite{Blanpied:2001ae}, Olmos de León \emph{et~al.}\cite{OlmosdeLeon:2001zn}, MacGibbon \emph{et~al.}\cite{MacGibbon:1995in} and Pasquini \emph{et~al.}\cite{Pasquini:2019nnx} are offset for clarity.}
\end{figure}
\par
\Fig{fig:chiEFT:compP} highlights a comparision of the chiral extrapolation prediction produced herein with a selection of recent experimental measurements. Excellent agreement is seen between the experimental measurements and the result obtained herein. This highlights the utility of the quark projection technique and partially quenched chiral effective field theory used herein. It validates our understanding of QCD behind their development and use.
\begin{figure}
  \includegraphics[width=\columnwidth]{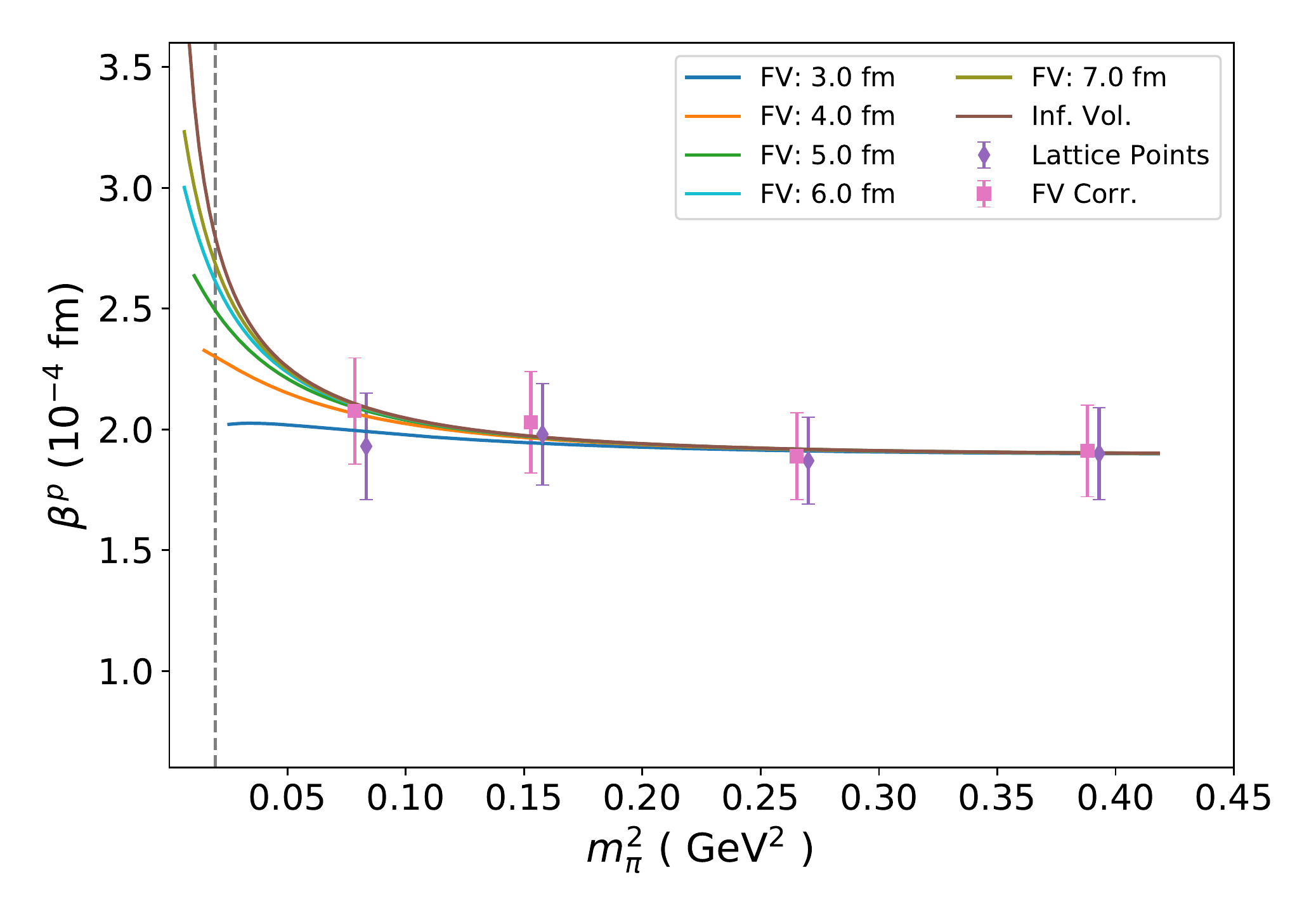}%
  \caption{\label{fig:chiEFT:FVP}Finite volume (FV) extrapolations of $\beta^{p}$ with total full QCD coefficients appropriate for fully dynamical background field lattice QCD simulations. The infinite-volume case relevant to experiment is also illustrated. Both finite-volume valence-valence lattice QCD results (Lattice Points) and their finite-volume (FV Corr.) total QCD corrected values are illustrated. The Lattice Points are offset for clarity.}
\end{figure}
\par
The chiral expansion of \eqnr{eqn:chiEFT:fitfunc} may also be used to guide future lattice QCD calculations at a range of lattice volumes by using the discretised sum forms of the continuum integrals with either valence-valence or total integral coefficients. \Fig{fig:chiEFT:FVP} shows finite volume extrapolations of $\beta^{p}$ with total full QCD coefficients for a range of lattice volumes, $3.0 \text{ fm}\leq L_{s} \leq 7.0$ fm and pion masses with $\mpi\,L_{s} \geq 2.4$. Here the $7.0$ fm result still differs from the infinite volume result by $\sim 4\%$.
\par
\begin{figure}
  \includegraphics[width=\columnwidth]{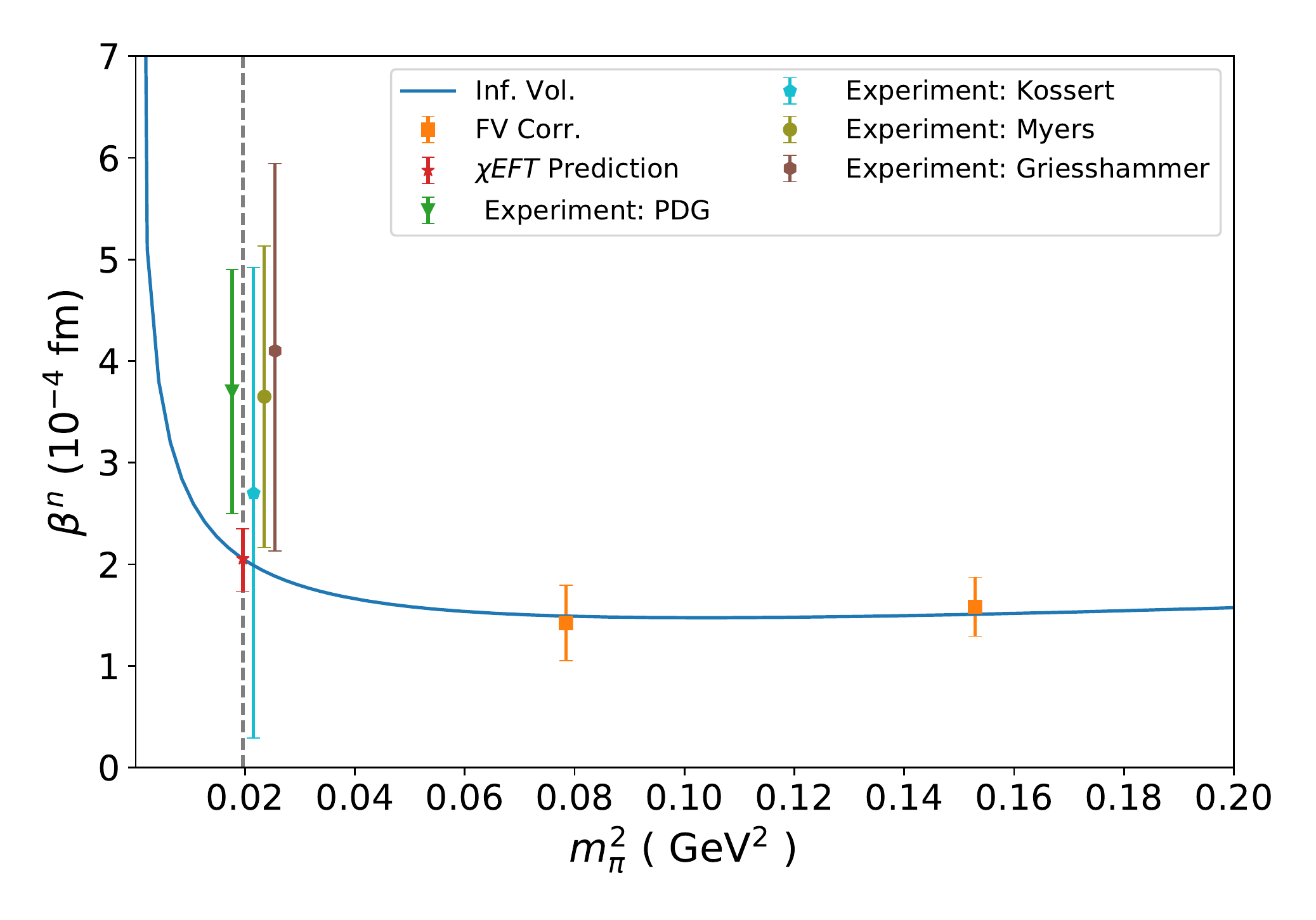}%
  \caption{\label{fig:chiEFT:compn}The magnetic polarisability of the neutron, $\beta^n$ from our chiral effective field analysis $\rb{\chi EFT\text{ Prediction}}$. Lattice results of this work, finite-volume corrected (FV Corr.) with total QCD coefficients are compared with experimental measurements via an infinite-volume (Inf. Vol) chiral extrapolation with total QCD coefficents. The error bar at the physical point reflects systematic and statistical uncertainties added in quadrature. Experimental results from Kossert \emph{et~al.}~\cite{Kossert:2002jc,Kossert:2002ws}, the PDG~\cite{Tanabashi:2018oca}, Myers \emph{et~al.}~\cite{Myers:2014ace} and Griesshammer \emph{et~al.}~\cite{Griesshammer:2012we} are offset for clarity.}
\end{figure}
The same process is used to predict the value for the magnetic polarisability of the neutron at the physical point to be
\begin{align}
  \beta^{n} = 2.06(26)\rb{{}^{+15}_{-20}} \times 10^{4}\text{ fm}^3,\nonumber
\end{align}
where the numbers in parentheses represent statistical and systematic errors respectively. This value is in very good agreement with the value obtained in our earlier work of \Refl{Bignell:2018acn} where the $U(1)$ Landau eigenmode projection technique is used with a chiral extrapolation to obtain $\beta^{n} = 2.05(25)(19) \times 10^{4}$ fm$^3$. This agreement indicates the success of both the $SU(3) \times U(1)$ and $U(1)$ eigenmode quark projection techniques. \Fig{fig:chiEFT:compn} presents a comparison of $\beta^{n}$ to recent experimental measurements where good agreement is also observed.
\begin{figure}
  \includegraphics[width=\columnwidth]{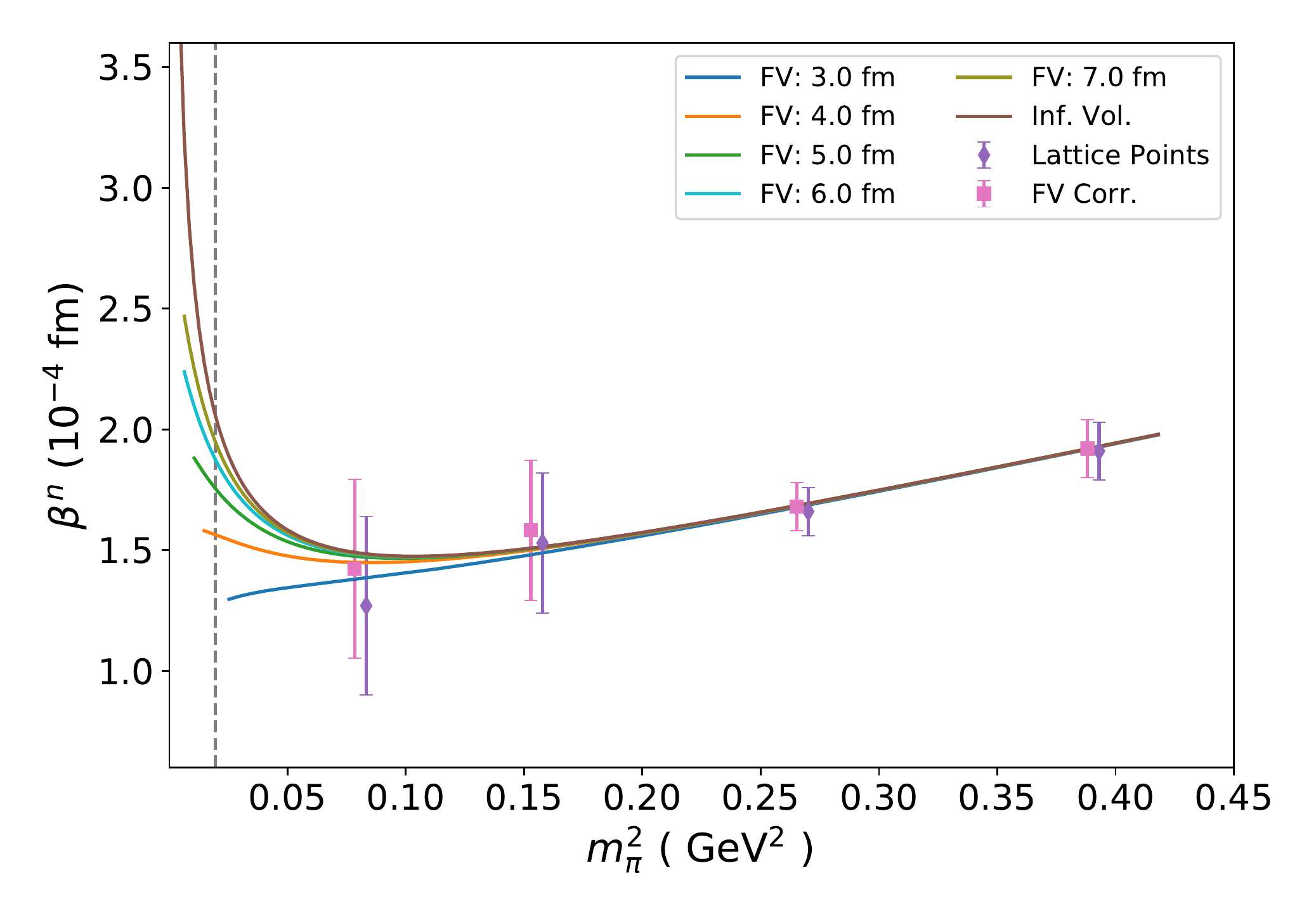}%
  \caption{\label{fig:chiEFT:FVn}Finite volume (FV) extrapolations of $\beta^{n}$ with total full QCD coefficients appropriate for fully dynamical background field lattice QCD simulations. The infinite-volume case relevant to experiment is also illustrated. Both finite-volume valence-valence lattice QCD results (Lattice Points) and their finite-volume (FV Corr.) total QCD corrected values are illustrated. The Lattice Points are offset for clarity.}
\end{figure}
\par
In an identical manner to the proton, \Fig{fig:chiEFT:FVn} presents finite-volume extrapolations with full QCD coefficients as a guide to future lattice QCD calculations. For both the proton and neutron, the leading and next-to-leading non-analytic terms of the chiral expansion give rise to a significant enhancement of the magnetic polarisabilities. In the case of the neutron, this chiral contribution reverses the trend observed on the lattice. Nevertheless, the coefficients of these non-analytic terms are model independent and well known. \Figtwo{fig:chiEFT:FVP}{fig:chiEFT:FVn} highlight the volume dependence of these contributions and indicate a significant challenge to directly observe these effects in future lattice QCD calculations. Here we note that the infinite volume value for the neutron is greater than that of the $L_{s} = 7.0$ fm lattice by $\sim 6\%$.
\par
As the lattice simulations are performed with only a single lattice spacing, it is not possible to directly quantify an uncertainty associated with taking the continuum limit. The non-perturbatively improved clover fermion action used herein has been shown \cite{Edwards:1997nh,Zanotti:2004dr} to display excellent scaling behaviour for hadron masses such that the $\order{a^2}$ corrections are expected to be small relative to the uncertainties already presented. Indeed \Refl{Edwards:1997nh} estimates a $<0.5\%$ error at the lattice spacing used in this study.
\par
Our conservative estimates for the systematic error due to the continuum limit extrapolation have a negligible effect on the final error when added in quadrature.
\subsection{Magnetic polarisability difference $\beta^{p} - \beta^{n}$}
The difference between the magnetic polarisability of the proton and the neutron can provide a test of Reggeon dominance~\cite{Gasser:2015dwa,Gasser:2020mzy}. When Reggeon dominance is assumed, the difference of magnetic polarisabilities can be predicted using chiral perturbation theory techniques and Baldin sum rules~\cite{BALDIN1960310,Gasser:2015dwa,Griesshammer:2012we}.
\par
\begin{figure}
  \includegraphics[width=\columnwidth]{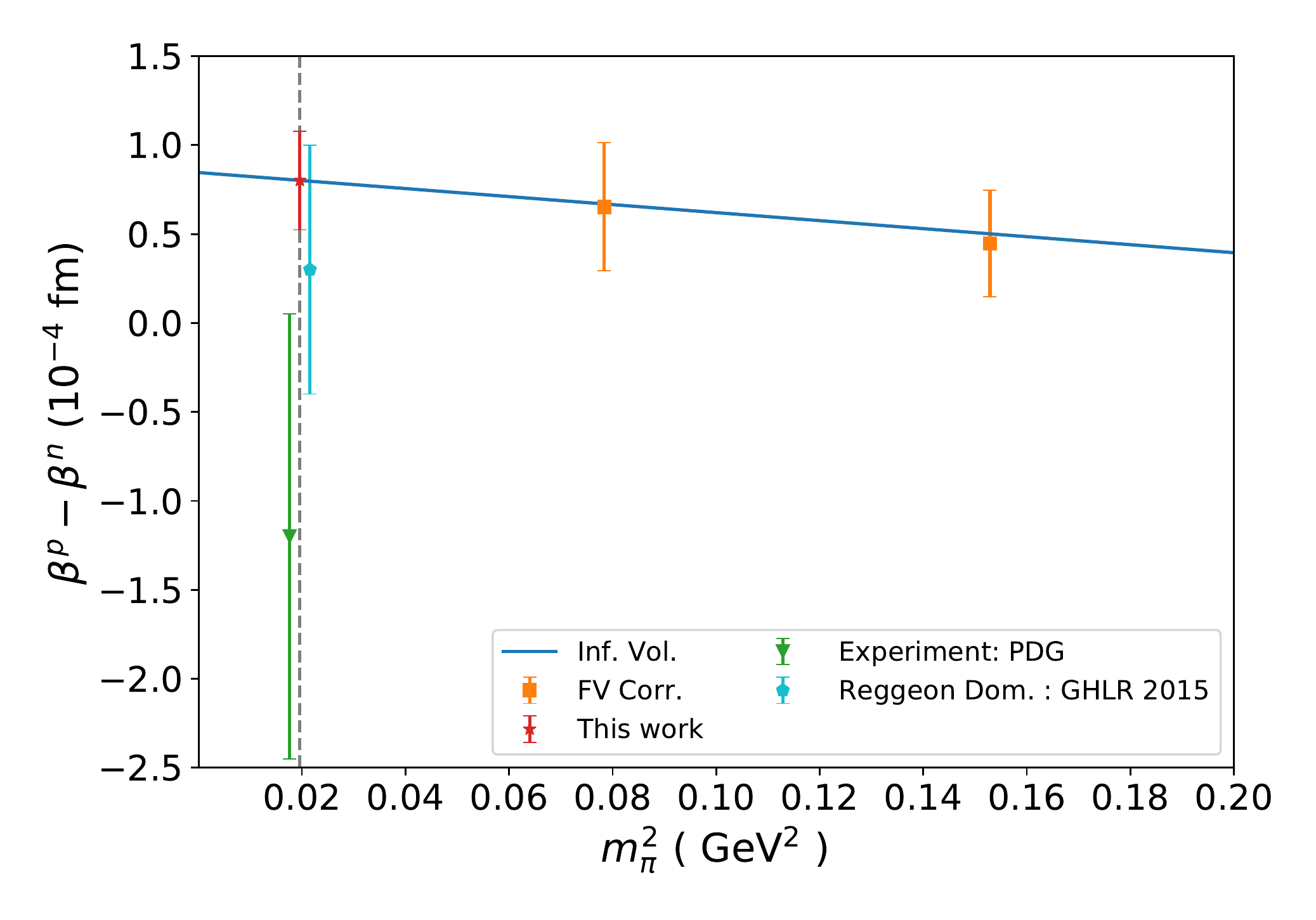}%
  \caption{\label{fig:chiEFT:pSUBn}The difference of the proton and neutron magnetic polarisabilities $\beta^{p} - \beta^{n}$. Lattice results of this work, finite-volume corrected (FV Corr.) with full QCD coefficients are compared with experimental measurements via an infinite-volume (Inf. Vol) chiral extrapolation. The error bar at the physical point (This Work) reflects systematic and statistical uncertainties added in quadrature. Experimental results from the PDG~\cite{Tanabashi:2018oca} and a Reggeon dominance prediction from Gasser \emph{et~al.}~\cite{Gasser:2015dwa} are offset for clarity.}
\end{figure}
We calculate the correlated difference between the magnetic polarisability of the proton and the neutron at each quark mass and then extrapolate to the physical regime using the formalism already discussed. In taking the difference, $u$-$d$ symmetry in the leading loop-integral coefficients of the chiral expansion in full QCD cause the contributions to cancel, leaving a simple linear extrapolation of our infinite-volume and full QCD corrected lattice results. The resulting prediction at the physical point is
\begin{align}
  \beta^{p} - \beta^{n} = 0.80\rb{28}\rb{4} \times 10^{4}\text{ fm}^3,\nonumber
\end{align}
where both statistical and systematic errors are indicated respectively. The central value differs from the difference between the extrapolated values discussed above due to the removal of rounding errors at each stage of the calculation. \Fig{fig:chiEFT:pSUBn} shows an extrapolation of all four lattice mass results to the physical regime with comparison to the PDG value~\cite{Tanabashi:2018oca} and a result derived using Reggeon dominance~\cite{Gasser:2015dwa}.

\section{Conclusions}
\label{sec:conc}
The magnetic polarisabilities of the proton and neutron have been calculated using asymmetric operators at the source and sink. Gauge invariant Gaussian smearing at the source encodes the dominant QCD dynamics while the \SUto eigenmode quark projection technique is used at the sink to encapsulate the low-lying quark-level Landau physics resulting from the presence of the uniform magnetic field.
\par
At the hadronic level, it is crucial to use a Landau wave function projection onto the proton two point correlation function as the proton is charged and hence experiences Landau level physics in a uniform magnetic field. The combination of these techniques has enabled constant plateau fits to be found in the magnetic polarisability energy shift of the proton for the first time.
\par
Furthermore, using the QCD gauge-covariant \SUto projection we are simultaneously able to produce magnetic polarisability energy shifts corresponding to both the neutron and proton ground states in a uniform background field. This represents a significant advance over the previous gauge-fixed $U(1)$ quark-level projection used to study the neutron polarisability.
\par
Connection with experimental results in the physical regime is achieved through the use of heavy-baryon chiral effective field theory and lattice QCD simulations at several pion masses. The resulting theoretical prediction for the magnetic polarisability of the proton is \hbox{$\beta^p = 2.79(22)\!\rb{^{+13}_{-18}} \times 10^{-4}$ fm$^3$} and \hbox{$\beta^n = 2.06(26)\!\rb{^{+15}_{-20}} \times 10^{-4}$ fm$^3$} for the neutron. These predictions are built upon \emph{ab initio} lattice QCD simulations using effective-field theory techniques to account for disconnected sea-quark-loop contributions, the finite volume of the periodic lattice and an extrapolation to the light quark masses of nature. These theoretical predictions are in good agreement with current experimental measurements and pose an interesting challenge for increased experimental precision.
\par
While we are necessarily in the confining phase of QCD, due to the small $\vB$ field strengths required for the perturbative energy expansion; from the success of the \SUto eigenmode projected quark sink technique it is clear that the external magnetic field has a significant effect on the distribution of the quarks within the nucleon.
\par
Our lattice results are electroquenched, they do not directly incorporate the sea-quark-loop contributions from the magnetic field. Future work would require a separate Monte Carlo ensemble for each value of $B$ considered and as such is prohibitively expensive due to a loss of QCD correlations. Another avenue that could be considered is to investigate the relativistic corrections to the energy-field expansion of \eqnr{eqn:EofB}. Here, improvements in lattice precsion will be required in order to succesfully fit the energies $\epm$ and construct the relativistic energy shift.
\par
It will be particularly interesting to extend this work to the case of hyperons. There the increased mass of the strange quark will illustrate differences between $\Sigmap$ and $p$ or $\Xiz$ and $n$ polarisabilities and give first insights into the environment sensitivity of quark-sector contributions to baryon magnetic polarisabilities. 



\begin{acknowledgments}
It is a pleasure to thank Heinrich Leutwyler for his comments highlighting how the proton and neutron magnetic polarisability difference provides a test of Reggeon dominance. We thank the PACS-CS Collaboration for making their $2+1$ flavour configurations available and the ongoing support of the International Lattice Data Grid (ILDG). This work was supported with supercomputing resources provided by the Phoenix HPC service at the University of Adelaide. This research was undertaken with the assistance of resources from the National Computational Infrastructure (NCI). NCI resources were provided through the National Computational Merit Allocation Scheme, supported by the Australian Government through Grants No.~LE190100021, LE160100051 and the University of Adelaide Partner Share.  R.B. was supported by an Australian Government Research Training Program Scholarship.  This research is supported by the Australian Research Council through Grants No.~DP140103067, DP150103164, DP190102215 (D.B.L) and DP190100297 (W.K).
\end{acknowledgments}

\bibliography{proton.bib}

\end{document}